\documentclass[%
 reprint,
superscriptaddress,
 amsmath,amssymb,
 aps,
pra,
]{revtex4-2}

\usepackage{graphicx}
\usepackage{dcolumn}
\usepackage[utf8]{inputenc}
\usepackage[colorlinks]{hyperref}
\usepackage{bm}
\usepackage{amsmath}
\usepackage{makecell}
\usepackage{multirow}
\usepackage{tabu}

\begin{document}

\title{Theoretical \textit{ab initio} Evolution of Satellite Intensity near Threshold for Cu K-shell transitions}

\author{Daniel Pinheiro}
\email{ds.pinheiro@campus.fct.unl.pt}
\affiliation{%
 LIBPhys-UNL, LA-REAL, Department of Physics, NOVA School of Science and Technology, NOVA University Lisbon, 2829-516 Caparica, Portugal.
}

\author{Gonçalo Baptista}
\affiliation{%
 Laboratoire Kastler Brossel, Sorbonne Université, CNRS, ENS-PSL Research University, Collège de France, Case 74; 4, place Jussieu, F-75005 Paris, France.
}

\author{César Godinho}
\affiliation{%
 LIBPhys-UNL, LA-REAL, Department of Physics, NOVA School of Science and Technology, NOVA University Lisbon, 2829-516 Caparica, Portugal.
}

\author{André Fernandes}
\affiliation{%
 LIBPhys-UNL, LA-REAL, Department of Physics, NOVA School of Science and Technology, NOVA University Lisbon, 2829-516 Caparica, Portugal.
}
\affiliation{
High Performance Computing Chair, Institute for Research and Advanced Training, University of Évora, Rua Romão Ramalho 59, 7000-671 Évora, Portugal.
}

\author{Jorge Machado}
\affiliation{%
 LIBPhys-UNL, LA-REAL, Department of Physics, NOVA School of Science and Technology, NOVA University Lisbon, 2829-516 Caparica, Portugal.
}
\affiliation{
High Performance Computing Chair, Institute for Research and Advanced Training, University of Évora, Rua Romão Ramalho 59, 7000-671 Évora, Portugal.
}

\author{Pedro Amaro}
\affiliation{%
 LIBPhys-UNL, LA-REAL, Department of Physics, NOVA School of Science and Technology, NOVA University Lisbon, 2829-516 Caparica, Portugal.
}
\affiliation{
High Performance Computing Chair, Institute for Research and Advanced Training, University of Évora, Rua Romão Ramalho 59, 7000-671 Évora, Portugal.
}

\author{Nancy Paul}
\affiliation{%
 Laboratoire Kastler Brossel, Sorbonne Université, CNRS, ENS-PSL Research University, Collège de France, Case 74; 4, place Jussieu, F-75005 Paris, France.
}

\author{Martino Trassinelli}
\affiliation{%
 Institut des NanoSciences de Paris, CNRS, Sorbonne Université, Paris, France.
}

\author{Miguel Avillez}
\affiliation{
High Performance Computing Chair, Institute for Research and Advanced Training, University of Évora, Rua Romão Ramalho 59, 7000-671 Évora, Portugal.
}
\affiliation{Department of Astronomy and Astrophysics, Technical University Berlin, Germany.}

\author{Paul Indelicato}
\affiliation{%
 Laboratoire Kastler Brossel, Sorbonne Université, CNRS, ENS-PSL Research University, Collège de France, Case 74; 4, place Jussieu, F-75005 Paris, France.
}

\author{José Paulo Santos}
\affiliation{%
 LIBPhys-UNL, LA-REAL, Department of Physics, NOVA School of Science and Technology, NOVA University Lisbon, 2829-516 Caparica, Portugal.
}

\author{Mauro Guerra}
\email{mguerra@fct.unl.pt}
\affiliation{%
 LIBPhys-UNL, LA-REAL, Department of Physics, NOVA School of Science and Technology, NOVA University Lisbon, 2829-516 Caparica, Portugal.
}
\affiliation{
High Performance Computing Chair, Institute for Research and Advanced Training, University of Évora, Rua Romão Ramalho 59, 7000-671 Évora, Portugal.
}







\begin{abstract}
In this work, we have investigated the evolution of satellite intensity near the ionization threshold for Cu K-shell transitions through theoretical methods. Employing standard state-of-the-art \textit{ab initio} methods, we have calculated all Cu K-shell transitions and simulated the full K$\alpha_1$ and K$\alpha_2$ spectrum where all transition parameters, as well as shake probabilities were determined theoretically. Through these calculations we show that standard state-of-the-art \textit{ab initio} methods achieve good agreement with experiment and enable us to simulate the intensity evolution near ionization thresholds within a good margin of error. Below-threshold satellite intensity was found to originate from resonant 1s$\rightarrow$3d and 1s$\rightarrow$4p excitations in Cu(I) and Cu(II) oxide phases respectively, which were included in our simulations.
\end{abstract}


\maketitle

\section{Introduction}
Theoretical atomic structure calculations of complex systems, such as multi-electron atoms and solids, rely on a variety of approximations, each associated with specific sources of uncertainty. In photon–atom interactions, one of the most widely used approaches is the sudden approximation (SA), in which the process responsible for modifying the effective potential experienced by the atomic electrons, such as ionization or nuclear decay, is assumed to occur on a timescale much shorter than that of electronic relaxation mechanisms, including shake-up and shake-off processes. In the case of ionization, this approximation implies that the time required for the ejected electron to leave the effective interaction region of the atomic potential is sufficiently short to be neglected. Under these conditions, the remaining electrons are assumed to respond instantaneously to the sudden change in the potential. Similar considerations apply to decay processes, which typically occur on sub-nanosecond timescales.

The sudden approximation provides an excellent description of high-energy and fast processes. However, its validity becomes questionable near ionization thresholds, where the outgoing electron has low kinetic energy and remains in the vicinity of the atomic system for a comparatively long time. As a consequence, the interaction time between the emitted electron and the residual ion increases, violating the fundamental assumptions underlying the SA. In this low-energy regime, the system approaches the adiabatic limit, in which the electronic structure evolves continuously during the transition. The gradual modification of the atomic potential allows the remaining electrons to partially adjust to the perturbation. This transition from the sudden to the adiabatic regime was previously investigated by Thomas \cite{thomas1984}, who employed time-dependent quantum mechanical methods to predict the relative intensities of satellite lines in core ionization processes.

Experimentally, this type of effects are most easily observed in x ray absorption spectroscopy (XAS). However, compared with emission spectroscopy, most XAS spectra do not have enough resolution to explore such effects in great detail. Some previous works have explored this using High Energy Resolution Fluorescence Detection (HERFD) \cite{Sier2024} and high resolution, reference-free double crystal spectrometers \cite{Ito2006, Ito2018, Ito2015, Ito2016, Mendenhall2017}.

In this work, we calculate and simulate, using an \textit{ab initio} approach, the satellite intensity of K-shell transitions in copper as a function of the beam's excitation energy. Throughout this text, the term "satellite intensity" will be used to refer to the total intensity derived from shake processes and will be used interchangeably with shake intensity. We expect to obtain a functional change in intensity near the threshold, comparable to the one obtained by Thomas \cite{thomas1984} and check it against experimental results. Given that the experimental data \cite{Galambosi2003} was obtained by x ray irradiation of a Cu foil, we expect that solid state effects will play a role, thus it is not expected that a perfect agreement with experimental spectra can be attained. Nonetheless, using only standard state-of-the-art theoretical calculations, we performed a simulation of the evolution of line intensity from adiabatic to sudden regime and obtained fairly good agreement with both experiment and the model proposed by Thomas.

\section{Theory}
\label{sec:theory}

For our theoretical calculations, we have used several computational methods to implement the equations and formalism described in this section. To accelerate our computations we have used distributed computing methods, such as GNU parallel, openMP and MPI. Some of the calculations have also been performed at the Oblivion supercomputer.

\subsection{Thomas Model}
\label{sec:theory_thomas}
Currently, one of the models used to predict the evolution of shake intensity, which can be observed as satellite line intensity in spectra, is the one by Thomas \cite{thomas1984}. This model was derived using time dependent quantum mechanics, building upon previous experiments performed by St{\"o}hr, Jaeger and Rehr \cite{stohr1983} as well as Carlson and Krause \cite{carlson1965}, where similar formulations were proposed. These experiments explored the intensities of shake-up and shake-off spectral lines respectively. 

In his work, Thomas aimed to predict the evolution from adiabatic to sudden regime and determine a model that was independent of the ionization mechanism. His goal was also to show that a time dependent approach with reasonable physical parameters was able to predict this evolution. By adding a time dependent potential, to approximately but explicitly include the interaction between the ejected electron and the remaining atomic system, the obtained model has the functional form of

\begin{equation}
    \mu = \mu_\infty \exp\!\Bigl(-\,\frac{r^2\,\Delta E^2}{15.32\,E_{\rm ex}}\Bigr),
    \label{eq:ThomasModel}
\end{equation}
\noindent
where $\mu_{\infty}$ is the satellite intensity in the sudden regime, $r$ is a distance comparable to the atomic system's dimensions in Angstrom (e.g. valence orbital mean radius), $\Delta E$ is the lowest transition energy above threshold and $E_{\rm ex}$ is the excess excitation energy above threshold, i.e. the ejected electron's kinetic energy. In practice, the variable $E_{\rm ex}$ is decomposed into the variable $E$ for the beam energy and the parameter $E_T$ for the threshold energy, where $E_{\rm ex} = E - E_T$. All energies are in electronVolt. The energy for the lowest excited shake-up transition is used for the value of $\Delta E$ as this is the first channel to open when the excitation energy is increased above threshold. Additionally, by expressing the satellite intensity as a ratio to the total radiative intensity, the total shake probability in the SA becomes a good approximation to the $\mu_{\infty}$. The total shake probability is only an approximation as it does not take into account the decay rates from each 1 and 2 holes configurations, which can result in changes to this ratio and is one of the factors we evaluate in this work. This model is employed by fitting the model to experimental data, varying the values of all parameters, starting from theoretical or previous experimental values.

\subsection{Roy Model}
A more recent model used to predict the evolution of shake intensity is the one by Roy \textit{et.al.} \cite{Roy2001}. This model builds upon the one determined by Thomas, giving the evolution of shake intensity as a function of the incoming photon energy, principal quantum number of the shake orbital and it's binding energy. This model has the functional form of
\begin{multline}
    P(\hbar\omega)=P(\infty)\left[\frac{\pi(2n-1)!}{2^{2n}(n-1)!(n+1)!}\right]^{-1}\\
    \times\int\limits_1^{E_p/E_B}\frac{(X-1)^{1/2}}{X^{n+2}\left\{1+\frac{1}{4}(2n+1)^2(E_B/E_p)X^2\right\}}dX,
    \label{eq:Roy}
\end{multline}
\noindent
where $\hbar\omega$ is the incoming photon energy and $E_p$ is the primary photoelectron energy, equal to $\hbar\omega-E_{\rm edge}$. $P(\infty)$ is the shake probability given by the sudden approximation, $n$ is the principal quantum number of the shake orbital and $E_B$ its binding energy.

\subsection{MCDF}
Similarly to previous works (e.g. \cite{Guerra2021, Pinheiro2022, Marques2020, Guerra2015}) we calculated all the transition energies, radiative and non-radiative rates, as well as shake probabilities in the SA regime using the state-of-the-art multiconfiguration Dirac-Fock (MCDF) method. To determine the atomic parameters needed, we write the atomic antisymmetric wave function $\psi$ as a linear combination of the  $\varphi$ configuration state functions (CSF): $\psi (1,2,\dots,N)=\sum_ia_i\varphi_i$, where $a_i$ are mixing coefficients. 

The relativistic general purpose multiconfiguration Dirac-Fock code (MCDFGME) developed by J. P. Desclaux, P. Indelicato and co-authors \cite{Desclaux1975, Indelicato1990, Indelicato2007, Santos2005} implements the MCDF method, using a self-consistent field approach. Various contributions, such as Coulomb and Breit (magnetic and retardation parts), both containing direct and exchange components, as well as quantum electrodynamics (QED) local potentials, such as vacuum polarization are included in the self-consistent field calculation. Other QED contributions, such as self-energy, are also included as perturbations to the final atomic wave function.

The optimized-level method was used to calculate the wave functions and energies of both the initial and final states, considering full relaxation, for the determination of the radiative rates. As each state was optimized separately, they are not necessarily orthogonal and, to calculate radiative rates, the formalism described by L{\"o}wdin \cite{Lowdin1955} was used. The length gauge was considered for all radiative transition rates.
For the non-radiative rates, the continuum-electron wavefunctions were obtained by solving the Dirac-Fock equations with the same atomic potential of the initial state. Here, no relaxation was allowed in both initial and final states.

In addition, the details of the adiabatic interaction near ionization threshold is not explicitly included in the calculations with the MCDFGME code. This interaction will be included through an intensity modulation mechanism explained in section \ref{sec:inten_modulation}.

\subsection{Synthetic Spectra}
To determine the satellite intensity we simulate full radiative spectra using the atomic calculations from MCDFGME. To do this we determined the energy, rate and width of all possible radiative and non-radiative subshell transitions. The total decay width of a configuration $i$ is given by the sum of partial widths $\Gamma_{ij}$ of all the possible decay paths from that particular configuration. In the case of radiative transitions the width is defined as

\begin{equation}
    \Gamma_i^R = \sum_j \Gamma_{ij}^R,
    \label{eq:radWidth}
\end{equation}
where $j$ represents all levels from configurations with lower energy than the $i^{th}$ subshell. For non-radiative transitions the width is defined similarly taking into account the creation of another hole in a subshell $k$

\begin{equation}
    \Gamma_i^{NR} = \sum_j \sum_k \Gamma_{ij;k}^{NR},
    \label{eq:radlessWidth}
\end{equation}
where $k$ represents a configuration with higher or equal quantum numbers than the subshell $j$. In the case of the $k$ subshell being equal to the $j$ subshell we have a Coster-Kronig transition, otherwise we have an Auger transition ($k > j$).

Using these subshell widths, we can simulate each transition's intensity in the experimental spectrum. For this we need to account for non-radiative decays, as well as the statistical weight $g_h$ of each decay path and normalize the intensity for each subshell. The spectral intensity of a transition $h$ is then determined as

\begin{equation}
    I_h = \frac{g_h}{\sum_d g_d} \frac{R_h}{R_i^{R} + R_i^{NR}},
    \label{eq:line_inten}
\end{equation}
where $h$ represents a decay path starting from the configuration $i$ and ending in configuration $j$. The statistical weights $g_d$ are summed over all decay paths $d$ starting from configuration $i$. $R_h$ is the calculated rate for the transition $h$. In addition to these quantities we make use of the shake probabilities explained in the next subsection to determine and simulate the intensities of diagram and satellite lines in our synthetic spectra \cite{Guerra2021, Guerra2015}.




\subsection{Total Shake Probability Calculation}
\label{sec:ShakeProbs}
To simulate the satellite intensity produced in our synthetic spectra, we need to first determine the probability of the atomic system undergoing a shake process. This probability can be calculated for each atomic orbital with the MCDFGME code, which can calculate the inner product between $n$ and $n-1$ electron atomic orbitals. This inner product is used to calculate the probability that an electron will stay bound in its orbital during an ionization. We also need to take into account the probability that an electron will change orbitals due to a correlated excitation process to another bound orbital. The total shake probability for a subshell can then be determined as

\begin{equation}
    P^{\rm shake}_{n\ell j} = 1 - \left[ \left| \int \Psi'^{*}_{n\ell j}(\vec{r}) \Psi_{n\ell j}(\vec{r})d\vec{r} \right|^2 \right]^{N_{n \ell j}} - P_0,
    \label{eq:TotalShake}
\end{equation}

where the integral corresponds to the inner product $\langle \Psi'^*_{n\ell j}|\Psi_{n\ell j} \rangle$ calculated using the MCDFGME code using the initial ($\Psi^*_{n\ell j}$) and final ($\Psi'^*_{n\ell j}$) states radial wavefunctions. $N_{n \ell j}$ is the number of electrons occupying the $n\ell j$ orbital. The parameter $P_0$ excludes the probability of transitions to other occupied bound states and can be written as

\begin{equation}
    P_0 = \sum_{n'} N_{n \ell j} \left| \int \Psi'^{*}_{n'\ell j}(\vec{r}) \Psi_{n\ell j} (\vec{r})d\vec{r} \right|^2,
    \label{eq:TransitionProb}
\end{equation}
the summation over $n'$ is performed across all occupied values of $n$ of the atomic system for which the transition cannot occur. For this probability, as the shake process within the Sudden Approximation (SA) \cite{Carlson1973} is a monopolar process, we only need to subtract the probability for transitions where the principal quantum number $n$ changes.

This probability has been, erroneously, interpreted in the past as the likelihood of one electron being excited into bound or continuum levels, but in fact this represents the probability of one \textbf{or more} electrons being shaken. In a recent paper by Nguyen \textit{et al.}, an expression for the calculation of a single electron shake was proposed using a simple binomial distribution \cite{Nguyen2022}

\begin{multline}
    P^{\rm single-shake}_{nlj} = \binom{N_{n \ell j}}{1} \left(1-(1-P^{\rm shake}_{n\ell j})^{\frac{1}{N_{n \ell j}}}\right) \times \\
    \times \left(1- P^{\rm shake}_{n\ell j}\right)^{\frac{N_{n \ell j}-1}{N_{n \ell j}}}.
    \label{eq:singleShake}    
\end{multline}

This expression then represents the probability of a single success for an electron not being found in the same subshell as it initially was. This is a more correct probability for the single shake process required for the \textit{ab initio} calculations. Multiple shake processes are much less likely to occur, but they can account for almost 10\% of the total shake probability for Cu \cite{Kochur2006} obtained with Eq. (\ref{eq:TotalShake}).

\subsection{Shake-off and Shake-up Probability Calculation}
\label{sec:SplitShakes}
In the case of the shake-up process, a bound electron is promoted to a higher $n$ orbital after photoionization. The probability of this process can be calculated directly using the inner product of the wavefunctions of the initial neutral configuration and the final ionized and excited configuration. Similarly to the probability used in equation (\ref{eq:TransitionProb}), the total shake-up probability for a subshell can be calculated as:

\begin{equation}
    P^{\rm shake-up}_{n\ell j} = \sum_{n'=n_1}^{\infty} N_{n \ell j} \left| \int \Psi'^{*}_{n'\ell j}(\vec{r}) \Psi_{n\ell j} (\vec{r})d\vec{r} \right|^2,
\end{equation}
where the summation is performed from the first available monopolar excitation to infinity for the total shake-up probability. To perform the summation to infinity we calculate the overlap for excitations up to $n=20$ and extrapolate for higher $n$ using a cubic bspline fit.

Using the total shake probability from Eq. (\ref{eq:TotalShake}), we can then separate the two shake processes, as:

\begin{equation}
    P^{\rm shake}_{n\ell j} = P^{\rm shake-off}_{n\ell j} + P^{\rm shake-up}_{n\ell j}.
    \label{eq:OFFplusUP}
\end{equation}


\subsection{Modeling the Energy Evolution of Line Intensity}
\label{sec:inten_modulation}
Although we are able to compute with high precision the atomic structure of multielectronic atoms as well as all the radiative and nonradiative decays, these results assume that there is enough energy to create all of the excited states that are included in the spectra simulation. This energy, which comes from the incident photon that will core-ionize the sample, determines the atomic states that can contribute to the spectral features which appear in a absorption or emission spectrum. In this context, "excitation" is used referring to the general process of promoting an atomic state of lower energy to a state of higher energy (ionization). Later in this article, "excitation" will also be used to refer to the resonant excitation process in specific Cu oxide phases.

Some of these effects can be included through the interaction cross section for the respective excitation mechanism. However, in particular for photoionization/excitation, this experimental information is scarce, in particular if we need to discriminate the formation cross sections for each atomic level. As the quantity of interest in this study is a ratio, if our cross sections are only regarding each atomic shell as in most photo-absorption studies, this contribution would cancel out. Theoretically, there have been some calculations for these cross sections \cite{Wang2025,Chernysheva2024,Dawra2022,Singh2019,Sardar2016,Liu1996,Fliflet1976}. However, these are often calculated with different theoretical frameworks and are not very accurate for the energy region near the threshold \cite{Sabbatucci2016, Brumboiu2019}. If we only consider a single energy dependent cross section for each atomic shell, as the transitions we are analyzing in this work result from decays of the K-shell, this would not impact our results ratios. In addition, the cross section value does not change the decay rate and shake probability calculations, i.e. this is always performed using the sudden approximation and is independent of the beam energy.

\begin{figure}[h]
    \centering
    \includegraphics[width=\linewidth]{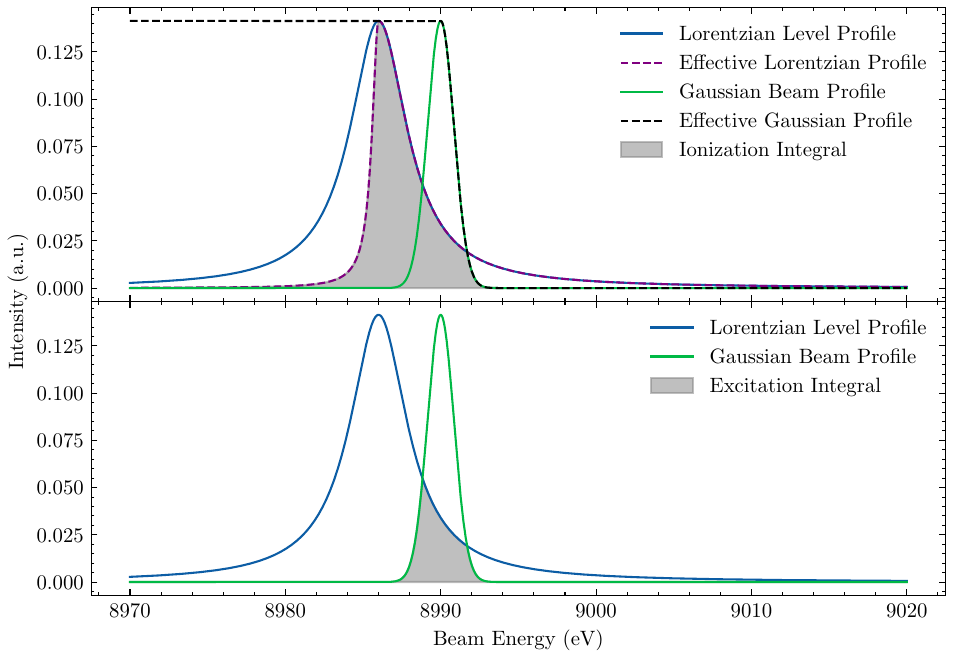}
    \caption{Graphical representation of the overlap integral between the beam Gaussian and initial level's Lorentzian profiles calculated for the modulation of line intensity. The representation for the ionization process is on the top and the excitation process on the bottom.}
    \label{fig:BeamOverlap}
\end{figure}

To describe the evolution of the shake probability, and as a result its intensity, as a function of the incident photon energy, the Thomas model (Eq. (\ref{eq:ThomasModel})) captures the phenomenology of the adiabatic to sudden transition via a simple exponential dependence on excitation energy. Our approach provides an equivalent, spectroscopy focused, perspective: by convolving the natural Lorentzian width of each ionized level with the Gaussian excitation profile of the beam, we directly model how the available excitation energy modulates line intensities near threshold. In order to consider all these effects, we designed a modulation function that provides a parsimonious representation of the system’s behavior, using sensible physical parameters. For this, the initial ionized level's width was used to modulate the intensity of each possible decay path through an overlap integral. Additionally, to consider other experimental sources of broadening, a Gaussian profile was also included for the excitation beam energy distribution. The final modulation of each decay is then calculated through the overlap integral between a Lorentzian profile with the level's width and the Gaussian profile for the beam energy distribution. In Fig. \ref{fig:BeamOverlap} we have a graphical representation of the calculated overlap integral. Such an integral will produce results similar to those used in recent studies, for example what can be seen in Figure 1 of \cite{Kavi2023}.

To account for the asymmetric nature of the ionization cross section, the Lorentzian profile used in the calculation of the integral was modified to quickly rise to maximum intensity near the ionization threshold (Effective Lorentzian Profile, Fig. \ref{fig:BeamOverlap} (top)). The modified profiles are described using a Heaviside step function:

\begin{multline}
    P_B(E; E_B, \sigma) = \\
        = ((1 - \theta(E - E_B)) + \theta(E - E_B)G(E; E_B, \sigma)),
\end{multline}
\begin{multline}
    P_L(E; E_B, \sigma, E_i, \Gamma_i^P) = \\
        = (1 - \theta(E - E_i))(2\ln{2}\sigma^2)L(E; E_i, \sigma) + \\
        + \theta(E - E_i)(\Gamma_i^P/2)^2L(E; E_i, \Gamma_i^P),
\end{multline}
where $G(E; E_B, \sigma)$ and $L(E; E_i, \Gamma_i^P)$ are regular Gaussian and Lorentzian profiles respectively, $E_B$ is the beam energy, $\sigma$ is the standard deviation of the beam's energy profile, $E_i$ is the target level's energy threshold and $\Gamma_i^P$ is the partial width (radiative and non-radiative) for the target level. The additional multipliers in the Lorentzian profile are used to normalize both profiles with each other when computing the integral and ensure there are no discontinuities at $E_B$ and $E_i$. The integral is then given by:

\begin{multline}
    F(E_B, \sigma_B, E_i, \Gamma_i^P) = \\
        = A\int_{-\infty}^{\infty} Min\left(P_B(E; E_B, \sigma)\right., \\
        \left.P_L(E; E_B, \sigma, E_i, \Gamma_i^P)\right) dE,
\end{multline}
and the total factor is then scaled by an amplitude $A = (2 / \Gamma_i^P\pi)$ to ensure that $\lim_{E_B\rightarrow\infty} F(E_B, \sigma, E_i, \Gamma_i^P) = 1$, i.e. the channel is fully open and we observe the SA results in the spectrum. The profiles used in the calculation for the excitation integral are not modified, i.e. $P_G(E; E_B, \sigma) = G(E; E_B, \sigma)$ and $P_L(E; E_B, \sigma, E_i, \Gamma_i^P) = (\Gamma_i^P/2)^2L(E; E_i, \Gamma_i^P)$.

In our simulations, we numerically solve both integrals using Simpson's rule, to accommodate arbitrary functions such as the piecewise functions for the ionization's process effective / final beam and level profiles (dashed lines in Fig. \ref{fig:BeamOverlap} (top)).

\section{Results}
\label{sec:Results}
\begin{figure}[h]
    \centering
    \includegraphics[width=\linewidth]{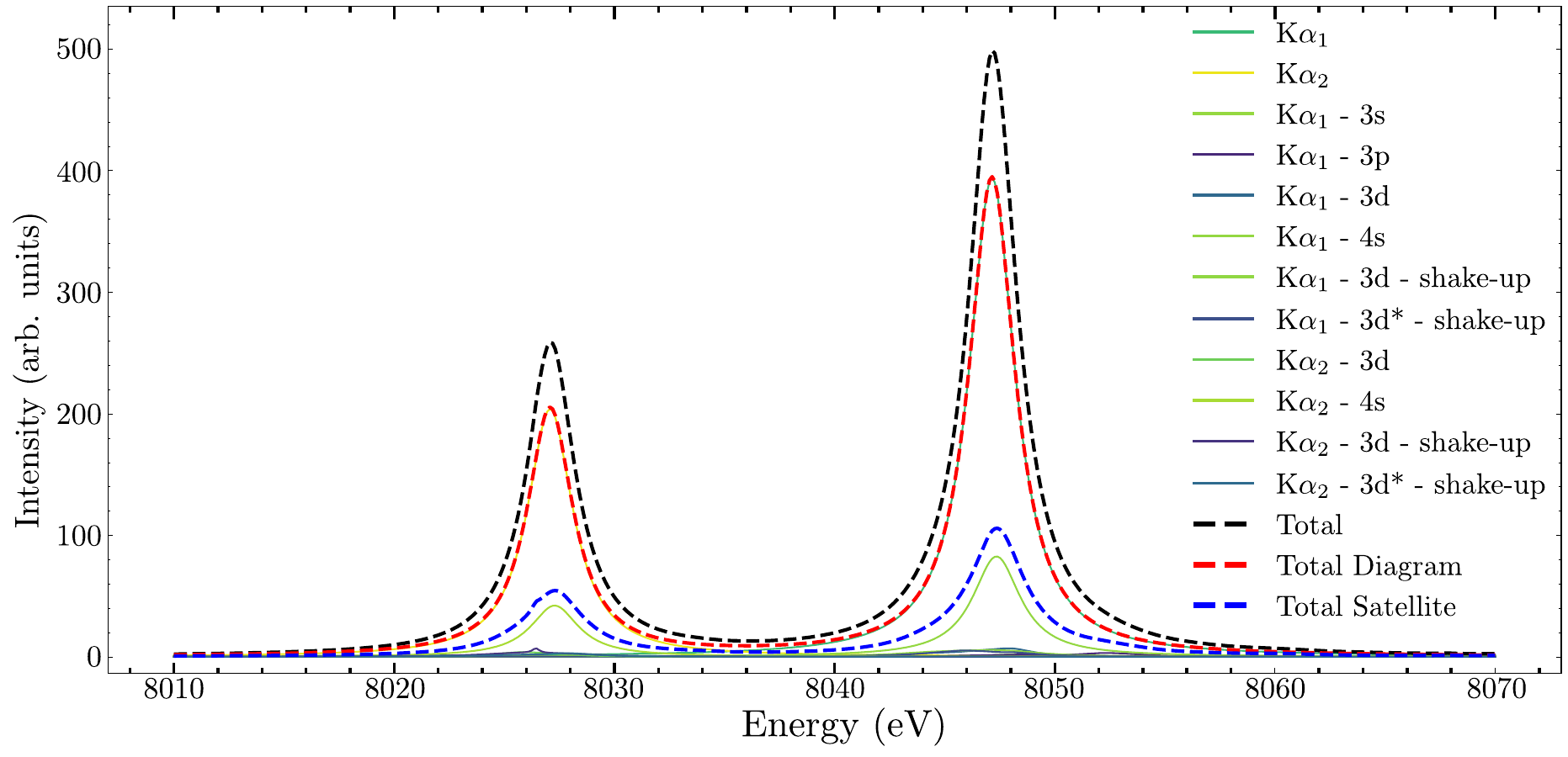}
    \caption{Simulated Cu K$\alpha_{1,2}$ synthetic spectrum. All satellite contributions have been included, however only the most intense were drawn for clarity. This simulation was performed for an infinite excitation energy.}
    \label{fig:CuTotalSpectrum}
\end{figure}

Using the results from atomic structure calculations performed with the MCDFGME code and the equations presented in section \ref{sec:theory}, we have calculated the synthetic spectrum for the Cu K$\alpha_{1,2}$ transitions. In the spectrum presented in Fig. \ref{fig:CuTotalSpectrum} we have included all satellite transitions weighted by the respective shake probabilities and ionization cross sections. Lower intensity transitions have been omitted for visual clarity.

For the shake calculation, using MCDFGME and following the calculations detailed in Sections \ref{sec:ShakeProbs} and \ref{sec:SplitShakes}, we first calculate the total shake for a 1s ionization, which was determined as $\approx25.968\%$. The calculated shake probability is then split into the probability of shake-off and shake-up processes, which were determined as $\approx13.080\%$ and  $\approx12.888\%$, respectively. Further decomposition of the total shake probabilities is shown in Table \ref{tab:ShakeProbs}. The 2J values shown correspond to the values of the total angular momentum of the Cu ground state. Additionally, only monopolar excitations were considered for the shake-up process. As the shake-up electron can in theory be excited to $n=\infty$, the total shake-up probability for all excitations was determined by extrapolation.

\begin{table}[h!]
    \centering
    \caption{Calculated Shake probabilities in \% for all levels used in the spectral simulation. The shake orbital is identified by its x-ray notation and the excited orbital by the atomic notation.}
    \label{tab:ShakeProbs}
    \setlength\tabcolsep{4.0pt}
    \begin{ruledtabular}
    \begin{tabular}{ccccc}
        \makecell{Shake \\ Orbital} & 2J & Shake-off & \makecell{Excited \\ Orbital} & Shake-up \\
        \hline
        K & 0 & 0.027 & 4s & 3.20$\times10^{-5}$ \\
           & 2 & 0.014 & 4s & 0.000 \\
        L$_1$ & 0 & 0.089 & 4s & 0.005 \\
           & 2 & 0.095 & 4s & 0.001 \\
        M$_1$ & 0 & 0.349 & 4s & 0.031 \\
           & 2 & 0.349 & 4s & 0.007 \\
        N$_1$ & 0 & 8.083 & 5s & 0.341 \\ 
           & 2 & 8.434 & 5s & 0.341 \\
        L$_2$ & 0 & 0.152 & 4p & 0.020 \\
           &   &       & 5p & 0.001 \\
           & 2 & 0.163 & 4p & 0.007 \\
           &   &       & 5p & 0.001 \\
        L$_3$ & 0 & 0.294 & 4p	& 0.026 \\
           &   &       & 5p & 0.006 \\
           & 2 & 0.305 & 4p & 0.017 \\
           &   &       & 5p & 0.005 \\
        M$_2$ & 0 & 0.661 & 4p & 0.116 \\
           &   &       & 5p & 0.019 \\
           & 2 & 0.677 & 4p & 0.155 \\
           &   &       & 5p & 0.040 \\
        M$_3$ & 0 & 0.883 & 4p & 0.441 \\
           &   &       & 5p & 0.108 \\
           & 2 & 0.761 & 4p & 0.404 \\
           &   &       & 5p & 0.179 \\
        M$_4$ & 0 & 0.168 & 4d & 3.560 \\
           &   &       & 5d & 1.080 \\
           & 2 & 0.374 & 4d & 2.260 \\
           &   &       & 5d & 1.140 \\
        M$_5$ & 0 & 2.272 & 4d & 2.490 \\
           &   &       & 5d & 1.500 \\
           & 2 & 1.943 & 4d & 3.690 \\ 
           &   &       & 5d & 1.810 \\
        \hline
        Sum& 0 &12.977 & Interpolated&12.779 \\
        Sum& 2 &13.114 & Interpolated&12.925 \\
        \multicolumn{2}{c}{Average} & 13.080 & & 12.888 \\
        \hline
        \multicolumn{2}{c}{Total} & \multicolumn{3}{c}{25.968} \\
    \end{tabular}
    \end{ruledtabular}
\end{table}

\begin{figure}[!h]
    \centering
    \includegraphics[width=\linewidth]{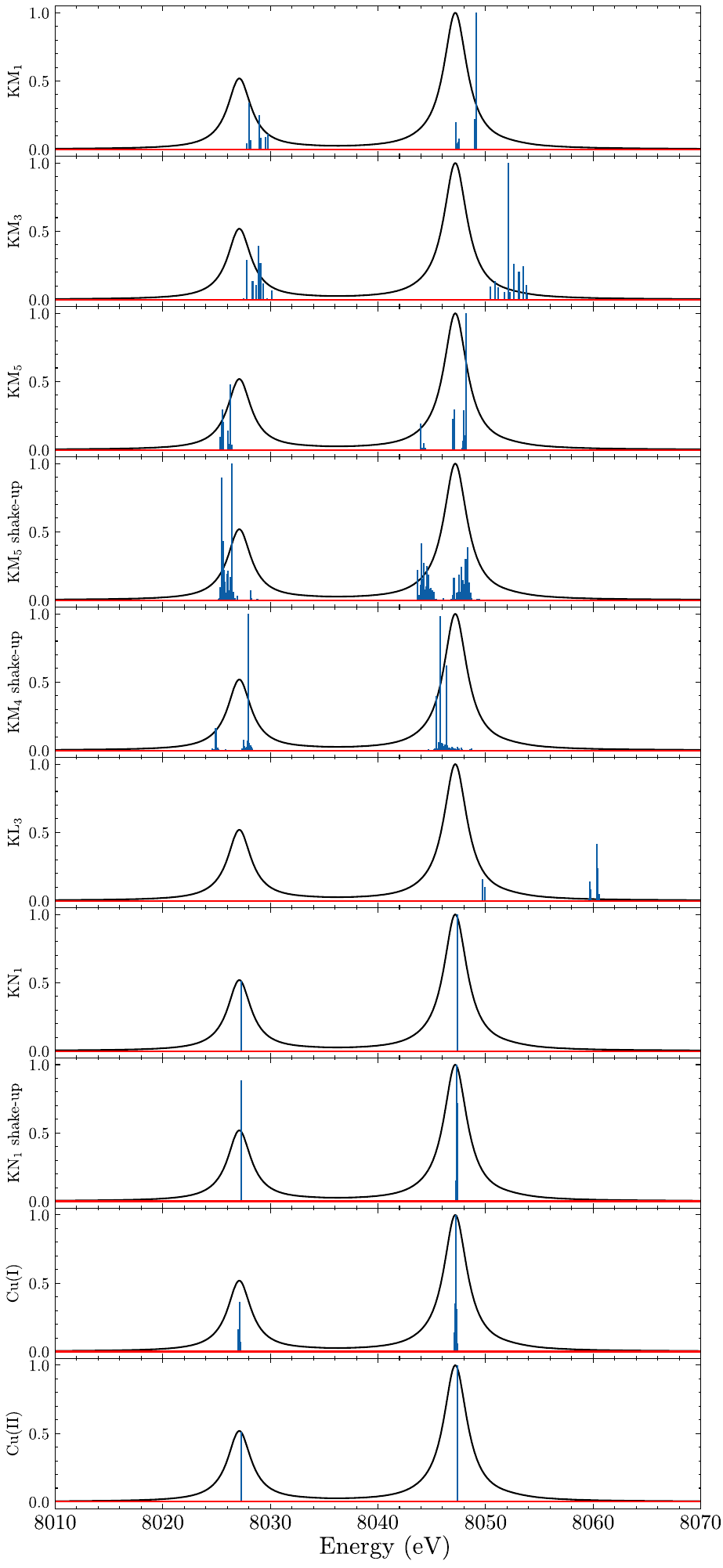}
    \caption{Stem plots or stick spectra of some of the satellite transitions included in the simulations. Transitions for the Cu(I) and Cu(II) ions are also included, which will be important later in this work. The Cu K$_{\alpha_{1,2}}$ spectra is also shown in black.}
    \label{fig:Cu_sticks}
\end{figure}

In Table \ref{tab:ShakeProbs} we only show the shake probabilities for lines that were included in the simulation, due to computation time and storage restrictions. In the case of the shake-off process this is not an issue, where we can see that all lines originating from this process were included. This is not the case for the shake-up process, as the number of combinations due to excitation is orders of magnitude larger. Comparing the total shake-up probability obtained from extrapolation ($12.779\%$ and $12.925\%$) and by summing the probabilities shown in Table \ref{tab:ShakeProbs} ($9.744\%$ and $10.057\%$) we see that most shake-up is present, however approximately $2.9\%$ is missing. To take this into account, before the simulation is performed, a routine is employed to determine how to distribute this missing probability into the existing shake-up decay channels. As excitations to higher $n$ orbitals would have closely spaced threshold energies, by distributing this missing probability into the existing lower $n$ excitations, we expect this to be a good approximation in return for saving computation time and storage space. With this approximation, our calculations took over 2 months to complete and occupied $\approx1$TB of disk space. We can observe some of this data, for the energy region of interest, as a set of stem plots or stick spectra in Fig. \ref{fig:Cu_sticks}. The Cu K$_{\alpha_{1,2}}$ spectrum is also shown in black for reference. These spectra also include transitions from the Cu(I) and Cu(II) ions that will become important later.

\begin{figure}[!h]
    \centering
    \includegraphics[width=\linewidth]{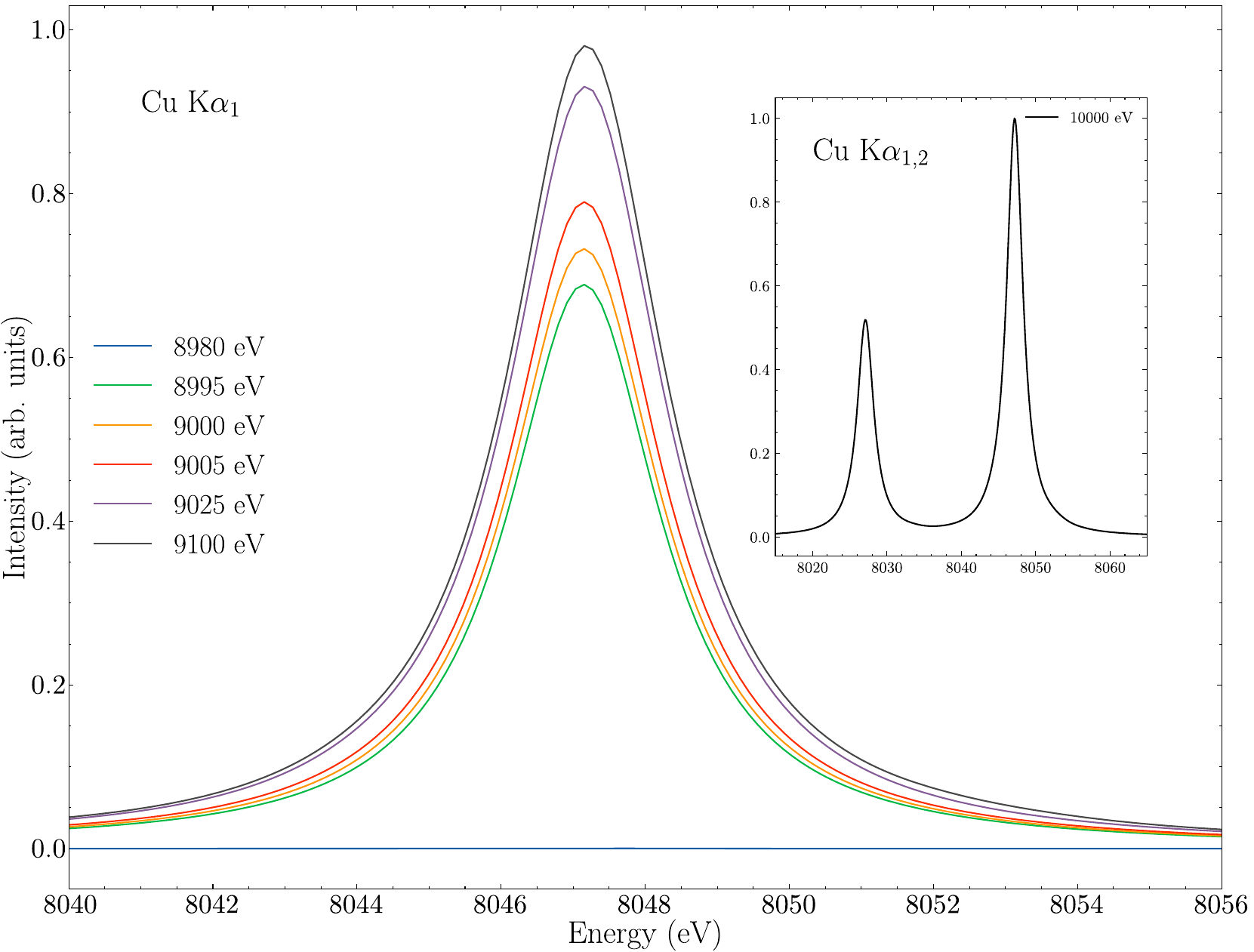}
    \caption{Evolution of the line shape of the simulated K shell transitions, with excitation energy. The energies were chosen as the ones shown in \cite{Galambosi2003}.}
    \label{fig:K_evolution}
\end{figure}

\begin{figure*}
    \centering
    \includegraphics[width=\linewidth]{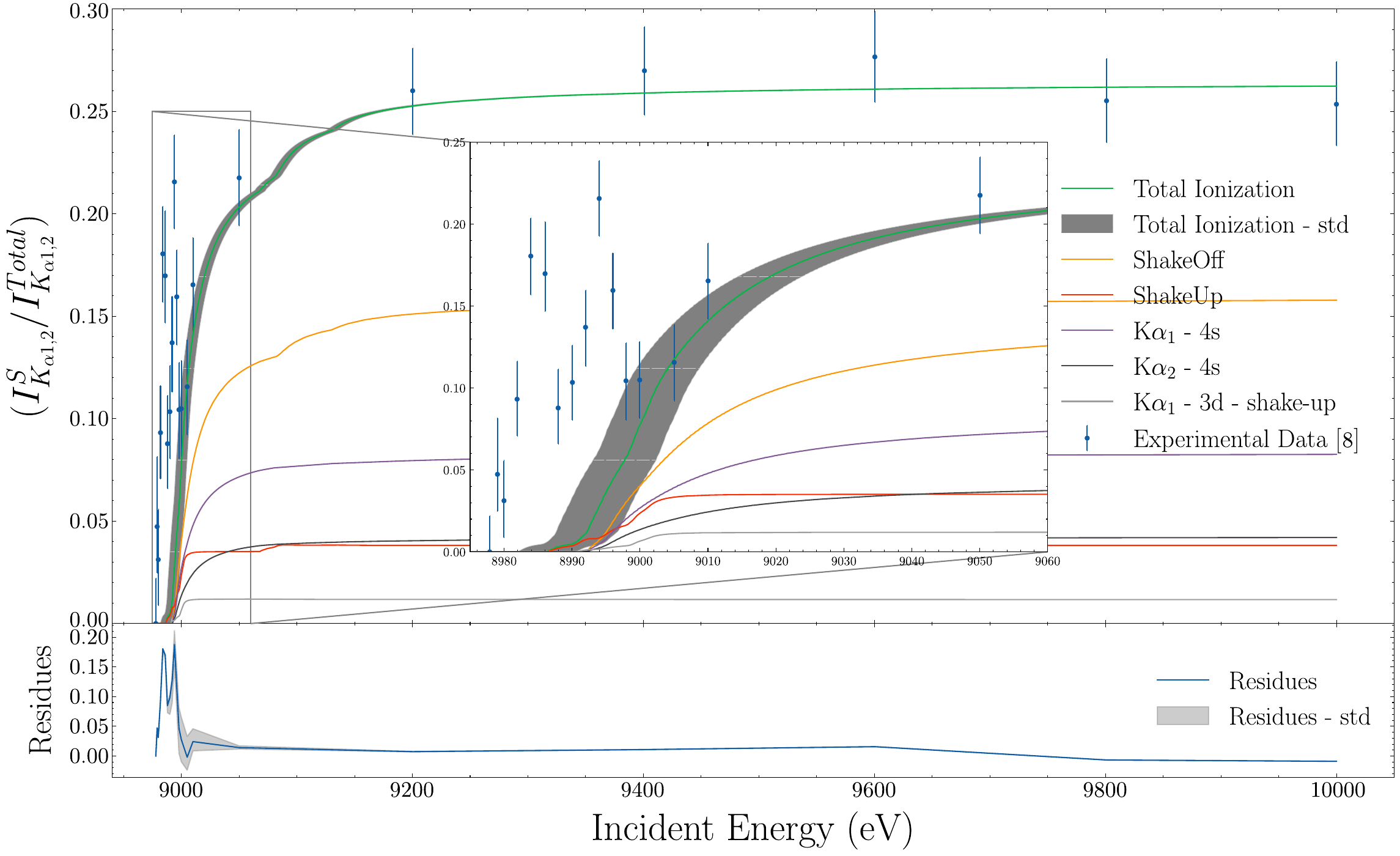}
    \caption{Evolution of the intensity ratio of the satellite lines, respective to the total K$\alpha_{1,2}$ intensity. The simulations were performed for experimental parameters of 0.2 eV for the beam width and 0.3 eV for the detector resolution, matching ref. \cite{Galambosi2003}. Only the most intense contributions are shown for visual clarity.}
    \label{fig:GalambosiRatios}
\end{figure*}

This data is then compiled in simulations such as the ones shown in Fig. \ref{fig:K_evolution}.
From such simulations we can then determine the relative intensity of satellite lines compared to the total intensity (Diagram + Satellites). The relative intensity can then be compared with the experimental results for each excitation energy. In Fig. \ref{fig:GalambosiRatios} we show the calculated ratios for the Cu K$\alpha_{1,2}$ satellites, as a function of excitation beam energy and compare it to the respective experimental data. With our simulations we can also further separate the contributions to this ratio, for each decay channel. The contributions from the most intense decay channels are also shown in Fig. \ref{fig:GalambosiRatios}. To make this comparison, an adjustment of the theoretical energy thresholds has to be made. This is due to the Fermi energy which arises from solid state effects and is not considered in the isolated atom model used in the atomic structure calculations.
For Cu the Fermi energy is around 7.0 eV. By shifting the energy for the K edge that we have obtained through our calculations ($\approx8986.0$ eV) by this 7.0 eV, we obtain $8979.0\pm0.5$ eV which is within the uncertainty interval of the energy determined experimentally ($8980.5\pm1.0$ eV \cite{NISTXRayDatabase}). However, for the satellite thresholds, we have two-hole configurations and hence the energy shift is harder to compute. A Bayesian analysis, nested sampling software called nested\_fit \cite{Trassinelli2019, Trassinelli2017} was used to fit the simulated profile to the experimental data, in order to determine the most likely energy shifts for each contribution. This leads to different offsets in the diagram and satellite lines' energies when compared to the isolated atom model used in our calculations. Further discussion of the results from fitting is done in Section \ref{sec:nested}.

The models proposed by Thomas and Roy \textit{et al.} were fitted to the simulation in Fig. \ref{fig:GalambosiRatios}, using only the ionization data, whereas the model proposed by Roy \textit{et al.} was also fitted to the experimental data. Both fits are not shown in the figure for visual clarity.

Starting with the model by Thomas, the fit to the simulation resulted in the parameter values: $\Delta E = 12\pm3$ eV; $r = 1.3\pm0.3$ \r{A}; $\mu_{\infty} = 0.26809\pm0.00005$; $E_T = 8986.87\pm0.05$ eV. The the same fit to the experimental data given in \cite{Galambosi2003} results in: $\Delta E = 10.2$ eV; $r = 1.1$ \r{A}; $\mu_{\infty} = 0.265$; $E_T = 8989.1$ eV. As no uncertainty was given in the experimental work, we assume that only the last significant digit is affected by uncertainty. Both $\Delta E$ and $r$ are in agreement, however for $\mu_{\infty}$ depending on the uncertainty of the experimental fit, this quantity might be in agreement with our calculations. If the uncertainty is higher than 1.2\% of the reported value ($0.003$) this is the case. The value of $E_T$, from the fit to the calculations, is approximately 2.2 eV lower than the one obtained experimentally, but is in very good agreement with the MCDF value for the first shake-up channel. As detailed in section \ref{sec:theory_thomas}, this parameter should correspond to the threshold energy of the lowest shake-up decay channel, which from our calculations is $E(KM_{4,5}\rightarrow4s) = 8987\pm4$. For the parameters in agreement, we can also compare their values with what we have theoretically calculated. For the radius, we have calculated the mean atomic spherical radius to be $r=1.047$, which also matches previous calculations \cite{Guerra2017}. The fitted value for this parameter shows a good agreement with the theoretical one. We also calculate the theoretical value for the parameter $\Delta E$, taking into account the energy shifts used for Figs. \ref{fig:GalambosiRatios} and \ref{fig:GalambosiRatios_exc} of $\Delta E=8\pm4$ eV. This is significantly lower than the fitted value to the simulated data and the one obtained from the experimental data, but still in agreement within the quoted uncertainties.

For the model by Roy \textit{et.al.}, we performed a fit using a linear combination of the models for the $n=3$ and $n=4$ shells, weighted by $c_1$ and $1-c_1$ respectively, as these are the orbitals that contribute the most to shake intensity. The fit to the experimental data resulted in the parameter values present in Table \ref{tab:RoyFit}.
\begin{table}[h!]
    \centering
    \caption{Fitting results for the linear combination of 2 Roy models for $n=4$ and $n=3$ shells. $E_{edge}$ and $E_B$ are given in eV.}
    \label{tab:RoyFit}
    \setlength\tabcolsep{1.0pt}
    \small
    \begin{ruledtabular}
    \begin{tabular}{ccccccc}
                & \multicolumn{2}{c}{Experiment} & \multicolumn{2}{c}{Simulation} & \multicolumn{2}{c}{Theory} \\
        \hline
                & $n=3$ & $n=4$ & $n=3$ & $n=4$ & $n=3$ & $n=4$ \\
        \hline
        $E_{edge}$ & \multicolumn{2}{c}{$8978.3(3)$} & \multicolumn{2}{c}{$8979.98(3)$} & \multicolumn{2}{c}{$8978.980(5)$} \\
        $P(\infty)$ & $0.27(3)$ & $0.28(3)$ & $0.28(2)$ & $0.28(2)$ & $0.26$ & $0.26$ \\
        $E_B$ & $9.9(3)$ & $5.0(3)$ & $13.15(2)$ & $4.91(4)$ & $13.100(5)$ & $6.660(5)$ \\
        \hline
        $c_1$ & \multicolumn{2}{c}{$0.28(3)$} & \multicolumn{2}{c}{$0.87(2)$} & \multicolumn{2}{c}{$0.46$} \\
    \end{tabular}
    \end{ruledtabular}
\end{table}
The values for both experiment and simulation do not agree within their quoted uncertainties, but they are reasonably close to the theoretical values. The fit values for both $P(\infty)$ are in agreement, but the remaining values are not. The binding energies for the experimental fit lie $24.4\%~(10.7\sigma)$ and $24.9\%~(5.5\sigma)$ from the theoretical values for $n=3$ and $n=4$ respectively and for the simulated data fit $0.4\%~(2.5\sigma)$ and $11.1\%~(43.75\sigma)$. The fit to the simulated data resulted in closer values for the binding energy parameters and a better representation of the contribution of $n=3$ and $n=4$ shake intensity through the $c_1$ parameter. These discrepancies show clear evidence that the experimental data contains more physical processes than ones considered in the model.

Looking at the simulated data we notice right away that, at lower energies, this data does not agree with the experimental one. For these energies, we do not predict any satellite intensity using only the ionization data. To reproduce what is experimentally observed in this region, we have explored experimental x ray absorption spectroscopy (XAS) (\cite{Sanson2021, Stern1975, Beccara2003, Fornasini2004, Gaur2012, Shimizu2001}) and more specifically x ray absorption near-edge spectroscopy (XANES) data for Cu (\cite{Klysubun2011, Alwis2015, Xu2023, Pankin2022}) where much more complete studies have been performed regarding the edge region and the identification of sample composition using resonant peaks in this edge region. These studies lead us to include excitations from Cu oxides, specifically the Cu(I) and Cu(II) K$\alpha_{1,2}$ transitions with a 4p and 3d spectator electron respectively as they were previously identified along the Cu K-edge at a similar energy to the resonant peaks we observe in the experimental data. The choice of these excitations from the XANES data will be explained further in Section \ref{sec:oxide_excitations}, where we explore the excitation processes in the edge region in more detail. Including these excitations in the simulation and assuming that they were treated as satellite intensity in the original data (their K$\alpha_{1,2}$ transitions fall in the same energy regions as the satellites with a 1s core-hole and a 3d or 4s spectator hole), we obtain ratios that agree with the data (Fig. \ref{fig:GalambosiRatios_exc}), for energies lower than 9060 eV.

The full simulation including these excitations can also be seen in Fig. \ref{fig:GalambosiRatios_full}, corresponding to the zoomed in region shown in Fig. \ref{fig:GalambosiRatios_exc}. As the excitations have a purely resonant nature, adding then to the simulations does not impact the intensity at higher energies. At higher energies, the decays from ionization of these oxides could be a contributing factor, however this would increase both total and satellite intensities which results in a similar ratio to bulk copper. Furthermore, the realistic amount of oxide on the surface of the sample could be around $\approx100nm$ resulting in a very low contribution from these species. For excitation this could still be observed as the excitation thresholds of the oxides are lower than the thresholds of the lowest copper satellites (see Table \ref{tab:thresholds}) and the cross section for excitation resonances can be orders of magnitude larger than for ionization \cite{Liu2014, Pradhan2025, Nahar2023}.

\section{Simulation Modeling}
\label{sec:Discussion}
From the calculated spectrum, we have determined the satellite intensity ratios for K$\alpha_{1,2}$ transitions. In Fig. \ref{fig:GalambosiRatios} and \ref{fig:GalambosiRatios_exc} the calculated ratios can be seen as a function of the beam energy, simulated using physical parameters such as detector resolution and excitation beam width. The values for these experimental parameters were used as reported in \cite{Galambosi2003}, i.e. a beam width (Gaussian profile) of 0.2 eV and a detector resolution of 0.3 eV. Table \ref{tab:simu_transitions} summarizes some examples of the initial and final LS coupled configurations of each type of transition included in the simulations. Even for the simulated spectral window from 8010 eV to 8070 eV, there are hundreds of satellite configurations, therefore only some examples are given. For the copper oxides, only the 4p and 3d excitations shown were considered (further details regarding these excitations in Section \ref{sec:oxide_excitations}).

\begin{table}[h!]
    \centering
    \caption{Examples of the initial and final configurations for each type of transition included in the simulations. As there are hundreds of satellite transitions in the simulated spectral window only some examples are given.}
    \label{tab:simu_transitions}
    \setlength\tabcolsep{1.0pt}
    \begin{ruledtabular}
    \begin{tabular}{cc}
        Type & Transition \\
        \hline
        Diagram & \makecell{1s 2s$^2$ 2p$^6$ 3s$^2$ 3p$^6$ 3d$^{10}$ 4s $\rightarrow$ \\ 1s$^2$ 2s$^2$ 2p$^5$ 3s$^2$ 3p$^6$ 3d$^{10}$ 4s} \\
        \makecell{Shake-Off \\ 3d (M$_{4,5}$)} & \makecell{1s 2s$^2$ 2p$^6$ 3s$^2$ 3p$^6$ 3d$^9$ 4s $\rightarrow$ \\ 1s$^2$ 2s$^2$ 2p$^5$ 3s$^2$ 3p$^6$ 3d$^9$ 4s} \\
        \makecell{Shake-Off \\ 4s (N$_1$)} & \makecell{1s 2s$^2$ 2p$^6$ 3s$^2$ 3p$^6$ 3d$^{10}$ $\rightarrow$ \\ 1s$^2$ 2s$^2$ 2p$^5$ 3s$^2$ 3p$^6$ 3d$^{10}$} \\
        \makecell{Shake-Up \\ 4s $\rightarrow$ 5s} & \makecell{1s 2s$^2$ 2p$^6$ 3s$^2$ 3p$^6$ 3d$^{10}$ 5s $\rightarrow$ \\ 1s$^2$ 2s$^2$ 2p$^5$ 3s$^2$ 3p$^6$ 3d$^{10}$ 5s} \\
        \makecell{Shake-Up \\ 3d $\rightarrow$ 4d} & \makecell{1s 2s$^2$ 2p$^6$ 3s$^2$ 3p$^6$ 3d$^9$ 4d $\rightarrow$ \\ 1s$^2$ 2s$^2$ 2p$^5$ 3s$^2$ 3p$^6$ 3d$^9$ 4d} \\
        \makecell{Cu(I) Excitation \\ 1s $\rightarrow$ 4p} & \makecell{1s 2s$^2$ 2p$^6$ 3s$^2$ 3p$^6$ 3d$^{10}$ 4p $\rightarrow$ \\ 1s$^2$ 2s$^2$ 2p$^5$ 3s$^2$ 3p$^6$ 3d$^{10}$ 4p} \\
        \makecell{Cu(II) Excitation \\ 1s $\rightarrow$ 3d} & \makecell{1s 2s$^2$ 2p$^6$ 3s$^2$ 3p$^6$ 3d$^{10}$ $\rightarrow$ \\ 1s$^2$ 2s$^2$ 2p$^5$ 3s$^2$ 3p$^6$ 3d$^{10}$} \\
    \end{tabular}
    \end{ruledtabular}
\end{table}

Due to the large uncertainties and the low number of points of the experimental data, we cannot make strong definitive conclusions; nonetheless, we will discuss the discrepancies and explore how this can be improved.

\subsection{Analysis Methods}
\label{sec:nested}
Before we begin discussing the results in detail, we should mention the adjustments made using the fitting procedure from nested\_fit used to determine the final simulated ratios presented in Fig. \ref{fig:GalambosiRatios} and \ref{fig:GalambosiRatios_exc}. The modulation of the ratios was obtained from the integrals described in Section \ref{sec:inten_modulation}.

For our initial fits, the direct use of these integrals and the inclusion of energy shifts for each process was not sufficient to reproduce the near threshold region of the experimental data. After further calculations we determined that, as the interactions during the excitation processes of diagram, shake-off, shake-up and excitation transitions processes are in principle different, the width of the initial atomic levels was not sufficient to uniquely describe their intensity evolution with excitation energy. To account for this, we have included a singular width multiplier for the level profile of each process. Each of these multipliers apply to all transitions of each decay process to prevent overfitting of the simulations to the experimental data, and guaranteeing that the resulting simulations stay grounded in physical data. In practice these multipliers adjust the edge and slope of the resulting modulation determined theoretically, to match the most likely features of the underlying physical process, by using a bayesian probabilistic approach. These factors, as well as the experimental energy shifts, were adjusted using nested\_fit. Furthermore, regarding the excitation data, as this intensity originates from different atomic structures (Cu(I) and Cu(II)) present in the sample in unknown amounts, and the excitation probabilities could not be normalized \textit{ab initio} to the ionization, a normalization multiplier to the intensity was required.

As the width multipliers had to be included in each simulation, these could not be directly included in the fitting procedure as parameters in the same manner that the energy shifts were. The simulation has to be performed for a set of multipliers, and only then can we fit to determine the best energy shifts and normalization multipliers. To determine the best set of multipliers using the nested\_fit program we make use of the Bayesian evidence determined for each fit, which is proportional to the model's probability. Each model corresponds to a choice of values for these width multipliers, determined through the simulation of excitation ($M_e$), diagram ($M_d$), shake-up ($M_u$) and shake-off ($M_o$) processes. The models are fitted to the experimental data by adjusting the remaining parameters described above, and the final evidence obtained in the fitting is then used to obtain the most likely set of $(M_d, M_o, M_u, M_e)$ width multipliers through interpolation.

The correlation between the multipliers is not precisely known in principle, therefore we would have to calculate a grid of simulations in a 4 dimensional space for our 4 width multipliers. However, we can assume that the excitation process should be independent from the ionization processes, allowing us to first determine the best excitation multiplier for each set of the 3 remaining multipliers $(M_d, M_o, M_u)$. This also allows us to reduce the dimensionality of our simulation grid to 3. To determine the parameter region where the maximum should be found, we first determined by hand a set of multipliers that resulted in a high evidence value. With this set as the center, we performed a number of simulations along each parameter axis, and through interpolation determined the best set of parameters.

\begin{figure}[h]
    \centering
    \includegraphics[width=\linewidth]{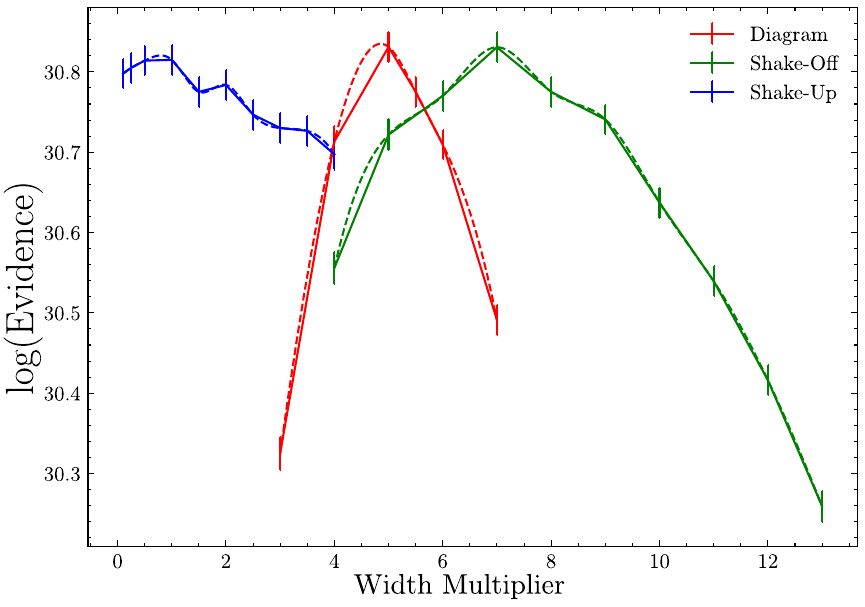}
    \caption{Set of simulations performed around the point (5.5, 8.0, 1.5) and their respective evidence. The maximum evidence was determined to be for the point (5.0, 7.0, 0.8).}
    \label{fig:find_middle}
\end{figure}

The evidence for each of these simulations is shown in Fig. \ref{fig:find_middle}, where we also show the interpolations used to determine the best parameter for each axis and the evidence range is reduced to better identify the maximum. Both diagram and shake-off processes show a clear maximum for their multipliers, with the shake-up process showing a more ambiguous maximum. This is attributed to two reasons, the first is the experimental data and energy region where the shake-up intensity rises. The shake-up process is responsible for the shake intensity that appears closest to the diagram edge, which results in a higher overlap with the excitation contribution from the Cu oxides near that region (8986 eV - 9000 eV). Due to the resonant nature of the excitation process, which is observed in the experimental data, the low number of points and high uncertainty are not sufficient to fully restrict the parameters for the shake-up process convolved with the excitation process near the threshold. This effect results in a much flatter evidence curve for the shake-up widths compared with the other 2 processes, where a clear maximum is visible and a $2\sigma$, corresponding to a 0.96 evidence difference can be determined around this maximum when we extend the range of multiplier values. In addition, the relative intensity of this process is $\approx25\%$ of the shake-off process, resulting in a lower overall contribution making it more difficult to separate from the total intensity.

The second reason is more related to the solid state physics where excited bound states result in bound electrons being promoted to the conduction band. This only happens in excitations or shake-up processes, resulting in a broad density of states which mix the various excitation and decay channels. This makes the threshold energies for each shake-up state ambiguous, as well as the width of the respective transitions and levels. On one hand, this makes the multiplier parameter for the shake-up process hard to determine with good accuracy, on the other it results in a better match between how the shake-up intensity rises in the experimental data and our calculations and model, which is represented by a multiplier $M_u$ closer to 1.

For our data, the initial maximum was determined to be for the set of multipliers (5.0, 7.0, 0.8). Although this is a reasonably fast way to find an approximate set of parameters that best fit the experimental data, it assumes that there is no correlation between them. The evidence for shake-up is only shown up to $M_u = 4$ for visual clarity, however it continues to slowly decrease with the same trend. This results in the shake-up width multiplier not having a pronounced effect on the final evidence of the fit, as opposed to the clear maxima of the other 2 multipliers. As such we also reduced the parameter space further, setting $M_u=1$ to match our theoretical calculations and evaluated the correlations between $M_d$ and $M_o$.

\begin{figure}[h]
    \centering
    \includegraphics[width=\linewidth]{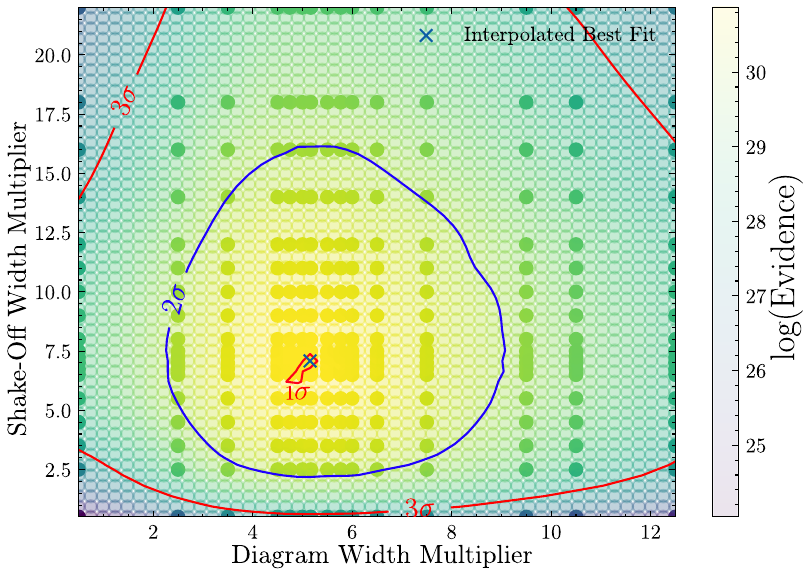}
    \caption{Visual representation of the final evidence interpolation performed to obtain the most likely model to describe the experimental data. The calculated grid points are represented with 100\% opacity and the interpolated data with 30\%. Contour lines corresponding to 1, 2 and 3$\sigma$ as also shown.}
    \label{fig:nested_cube}
\end{figure}

To fully consider the correlation between multipliers, a full 2 dimensional grid of simulations has to be calculated and interpolated, as we have done in Fig.\ref{fig:nested_cube}, around the point (5.0, 7.0), with $M_u=1$.
To obtain this final simulation grid, an initial grid of points was calculated with a step of 0.5 in each axis. To refine this grid, a new set of points was calculated in such a way that a new line perpendicular to each axis was added at the interpolation maximum. This was performed iteratively until the highest evidence fit was found near the highest evidence point.

From the interpolation we determine the simulation that most likely describes the experimental data, with the respective multipliers and remaining fit parameters. Furthermore, we can also see that both multipliers are not correlated as the evidence contours are approximately circular. This means that the two decay processes are not interfering with each other at a significant rate. In principle the shake and diagram processes are anti-correlated as the total decay has to add up to 1. However, these multipliers are only applied after the appropriate distribution of events between the two processes, removing the inherent correlation. The uncertainties from the fit parameters can be determined directly from nested\_fit, however the uncertainties of the width multipliers have to be obtained from interpolation using the evidences. Taking the maximum evidence point from the interpolation $\approx(5.153, 7.082)$, we perform a set of simulations by changing the multipliers independently along each of the axis, as the initial set in Fig. \ref{fig:find_middle}.
\begin{figure}[h]
    \centering
    \includegraphics[width=\linewidth]{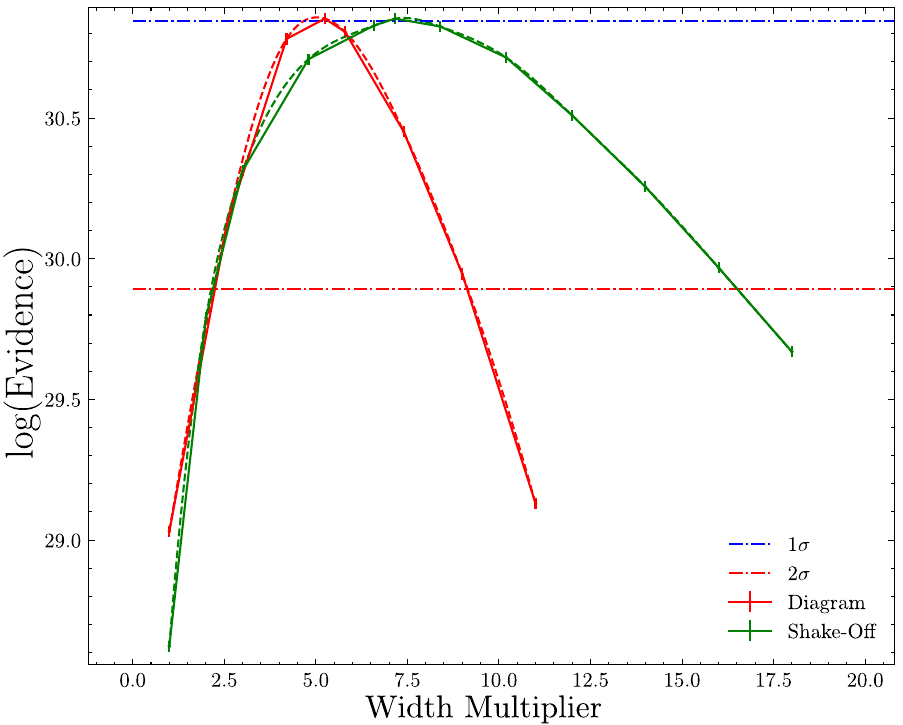}
    \caption{Final evidence curves used to determine the uncertainty of the width multipliers. Horizontal lines for $1\sigma$ and $2\sigma$ are also drawn for reference.}
    \label{fig:finalUncs}
\end{figure}
From these curves (Fig. \ref{fig:finalUncs}) we can now determine the uncertainties by calculating the values at a distance of $2\sigma$, corresponding to an evidence difference of $0.96$ relative to the maximum evidence.

\begin{table}[h!]
    \centering
    \caption{Fitting results obtained from nested\_fit. The intensity of lines from Cu(I) and Cu(II) excitations were multiplied by a factor to be normalized to the ionization data. Cu(II) lines were also further multiplied by a fraction relative to the Cu(I) intensity.}
    \label{tab:FitResults}
    \setlength\tabcolsep{4.0pt}
    \begin{ruledtabular}
    \begin{tabular}{ccccccc}
        \multicolumn{2}{c}{} & \makecell{Width \\ Multiplier} & \makecell{Energy \\ Shift} & Intensity \\
        \hline
        \multicolumn{2}{c}{Diagram} & $5\pm_{3}^{4}$ & $-7.0\pm0.6$ & \multirow{3}{*}{N.A.} \\
        \multicolumn{2}{c}{Shake-off} & $7\pm_{5}^{10}$ & $-10\pm5$ & \\
        \multicolumn{2}{c}{Shake-up} & $1.0$ & $-7\pm4$ & \\
        \hline
        \multirow{3}{*}{Exc.} & A & \multirow{3}{*}{0.54$\pm_{0.02}^{0.03}$} & $-1\pm2$ & $0.14\pm0.08$ \\
        & B &  & $-1\pm1$ & \multirow{2}{*}{$13\pm3$} \\
        & C &  & $9\pm1$ & \\
    \end{tabular}
    \end{ruledtabular}
\end{table}

\begin{figure*}
    \centering
    \includegraphics[width=\linewidth]{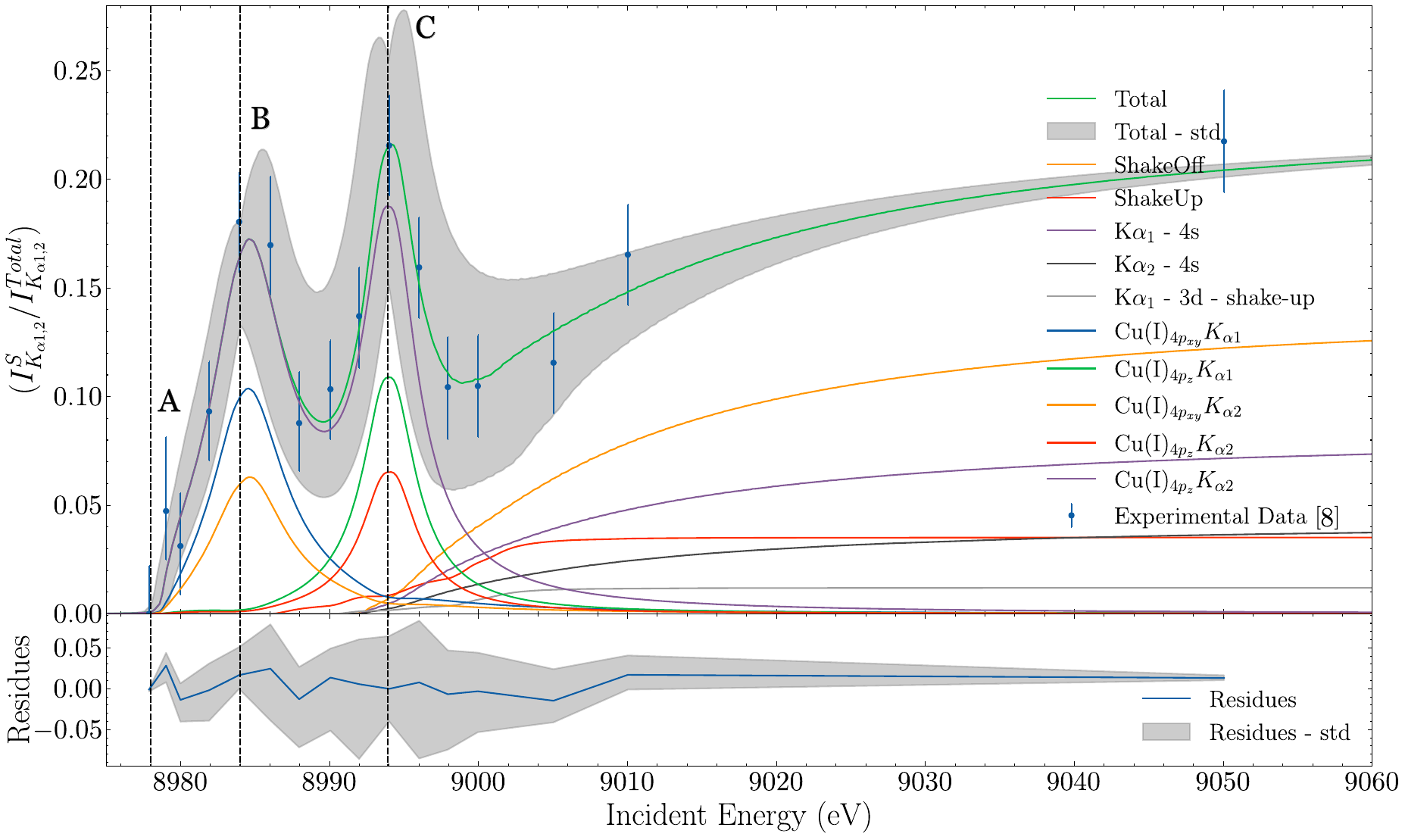}
    \caption{Evolution of the intensity ratio of the satellite lines, respective to the total K$\alpha_{1,2}$ intensity for the region with incident energy lower than 9060 eV. The intensity of excitations from Cu(I) and Cu(II) were added to the satellite intensity, resulting in 3 resonances labeled with A, B and C (using the fitted energy values in Table \ref{tab:thresholds}). Only the most intense contributions are shown for visual clarity.}
    \label{fig:GalambosiRatios_exc}
\end{figure*}
\begin{figure*}
    \centering
    \includegraphics[width=\linewidth]{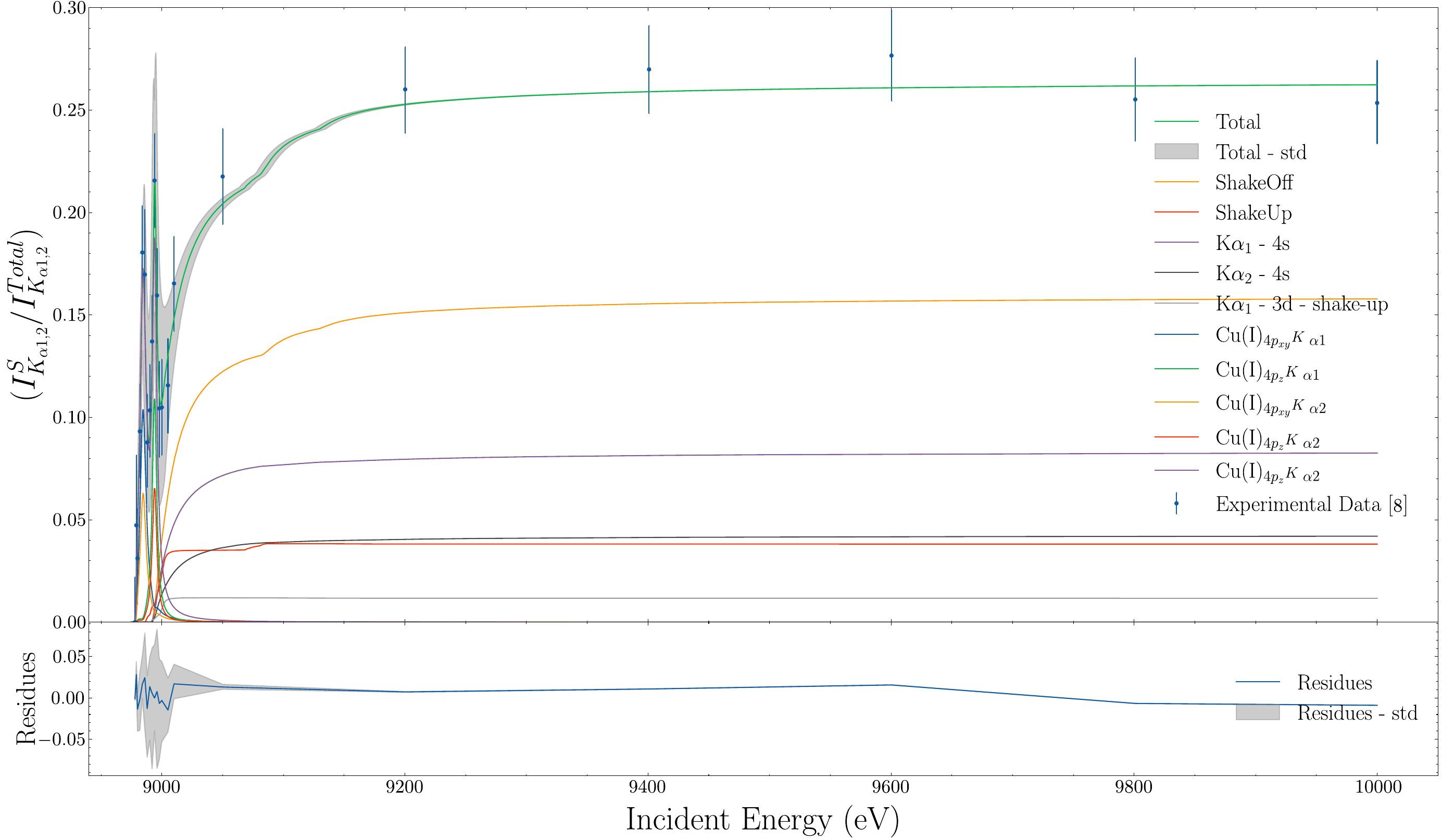}
    \caption{Evolution of the intensity ratio of the satellite lines, respective to the total K$\alpha_{1,2}$ intensity. The intensity of excitations from Cu(I) and Cu(II) were added to the satellite intensity. Only the most intense contributions are shown for visual clarity. Fig. \ref{fig:GalambosiRatios_exc} shows the excitation region near the threshold in more detail.}
    \label{fig:GalambosiRatios_full}
\end{figure*}

In Table \ref{tab:FitResults} we show the width multipliers for each excitation type, as well as the energy thresholds determined from the energy shifts. In addition, we also show the intensity multipliers for the excitation contributions which resulted in a very low value for the A peak and a very high value for the B and C peaks. With the most likely values determined for each of the parameters we can now calculate the final simulated intensities as shown in Figs. \ref{fig:GalambosiRatios_exc} and \ref{fig:GalambosiRatios_full}. The excitation region is shown in more detail in Fig. \ref{fig:GalambosiRatios_exc} and the simulation for the full range of experimental points is shown in Fig. \ref{fig:GalambosiRatios_full}.

For the A peak this is due to the contribution being in an area with only 3 experimental points, not providing enough statistic and resulting in the most likely case determined by nested\_fit being when this contribution practically disappears. For the B and C peaks, the high value can be due to two different reasons, one is the modulating function used not being correctly normalized in this case, as the maximum value of the integral depends on both the level's partial width and the width of the beam's energy distribution. The second reason is the excitation cross section can have a much higher value at the specific resonant energy compared to the remaining ionization data \cite{Liu2014, Pradhan2025, Nahar2023}. From the values of the fitted energy shifts and the energy levels that we have calculated theoretically we can now determine the experimental thresholds for each of the levels. The lowest energy thresholds determined, i.e. nearest to the edge, for both ionization and excitation can be seen in Table \ref{tab:thresholds}.

\begin{table}[h!]
    \centering
    \caption{Lowest energy thresholds for the simulated levels, for both ionization and excitation, determined from theoretical atomic calculation and nested\_fit.}
    \label{tab:thresholds}
    \setlength\tabcolsep{2.0pt}
    \begin{ruledtabular}
    \begin{tabular}{c|ccc}
        & Level & \makecell{Fitted Energy \\ Threshold} & MCDF \\
        \hline
        \multirow{3}{*}{Cu} & K & $8978.9\pm0.6$ & $8985.9$ \\
        & KN$_1$ & $8992\pm5$ & $9002$ \\
        & KM$_{4,5}$ & $9001\pm5$ & $9010$ \\
        & KM$_{4,5}$$\rightarrow$4s & $8987\pm4$ & $8994$ \\
        \hline
        Cu(II) & K$\rightarrow$M$_{4,5}$ (A) & $8978\pm2$ & $8979$ \\
        \multirow{2}{*}{Cu(I)} & K$\rightarrow$N$_{2,3}$(xy) (B) & $8984\pm1$ & \multirow{2}{*}{$8985$} \\
                               & K$\rightarrow$N$_{2,3}$(z) (C) & $8994\pm1$ &  \\
    \end{tabular}
    \end{ruledtabular}
\end{table}

From these results, we observe a value for the K level that agrees with the experimental value of $8980.5\pm1.0$ eV \cite{NISTXRayDatabase}. The remaining level's thresholds cannot be compared to experiments, as to our knowledge no experimental measurement has been done for these levels, only being able to be compared to theoretical calculations. The uncertainties of the energy thresholds for the shake levels are rather large compared to the one for the diagram level, as these shake contributions are much smaller components of the spectrum which are also convolved with the remaining intensity. These uncertainties, compared with the spectrometer resolutions, do not allow us to make strong conclusions about these thresholds.

\section{Discussion}
Now that we have detailed the simulation and analysis methods, we can discuss in more detail each of the regions and phenomena observed in the results we obtained.

\subsection{Ionization Data and Shake Probabilities}
Starting with the results from Fig. \ref{fig:GalambosiRatios}, the simulated ratios appear to be in good agreement with the ones obtained from experiment. These results were obtained using only ionization data, which means that the region between $8975$ eV and $\approx8998$ eV is not well represented. The increase in ratio from $\approx8990$ eV to $9050$ eV is in good agreement with the data from experiment. Above $9050$ eV we observe a fluctuating evolution of the ratios obtained from experiment. This could be due to solid state effects, such as the photoelectron scattering observed in Extended x ray absorption fine structure (EXAFS) spectroscopy (\cite{Sanson2021, Stern1975, Beccara2003, Fornasini2004}) or due to missing transitions, in the energy region of interest, from excitation of other elements or Cu species present in the sample. Nonetheless, due to the large uncertainty in the experimental data, our simulated ratios agree with the experimental ones for the whole region above 9200 eV, where all decay channels can be considered to be open.


At 10 keV, where the experimental ratio appears to have stabilized, we obtain a simulated value of $\mu=0.26238\pm_{0.00002}^{0.00004}$. The uncertainty for this value is very small, corresponding to mostly statistical uncertainty from the fit, as no uncertainty was given to the simulated intensities and theoretical decay rates. As mentioned previously, this value can be approximated by the total shake probability in the SA, which we have calculated to be 0.260 (Section \ref{sec:Results}). The total shake probability is only an approximation as it does not take into account the differences of decay rates for the 1 and 2 hole configurations, only the total formation ratio of 2 hole configurations. Nonetheless, this shows that the total shake probability in the SA is a very good approximation to this ratio. In addition, both values are well within the experimental uncertainty given in \cite{Galambosi2003} for the point at 10 keV of $0.25\pm0.02$, with both ratios being higher than the experimental value. The physical quantity that mainly affects this value in our simulations is the probability of the shake processes. We calculate these values from the atomic wavefunctions, but experimentally there is not a very precise agreement for the value of the shake probability in Cu. Although this value has previously been determined through various methods, both experimentally and theoretically, the values range from $\approx23\%$ to $\approx38\%$ \cite{Chantler2010, Deutsch1995, Ito2006, Sauder1977, Mukoyama1987}. The most commonly reported values range from $\approx26\%$ for theoretical calculations and $\approx30\%$ from experimental analysis. This puts our value in agreement with previous works, although lower than the experimental values.

\subsection{Excitation region}
\label{sec:oxide_excitations}
Now that we understand the simulations for the ionization data, we can focus on the region between $8975$ eV and $9050$ eV, where the ionization data is not sufficient to describe all the experimental results. In this region, XANES effects could dominate, as collective excitations to final states might result in slightly shifted K$\alpha_{1,2}$ lines. Although we are performing x ray fluorescence of Cu in the solid, instead of absorption, the underlying physical processes that change fluorescence intensity near the ionization edge still occur in the sample and should be included in the model to describe the experimental data.

In both theory and experimental literature, we observed absorption spectra with a similar structure of three peaks along the rising edge \cite{Gaur2012, Shimizu2001, Klysubun2011, Alwis2015, Xu2023, Pankin2022}. From the XANES data, these peaks are usually identified as A, B and C, corresponding to excitations of Cu oxides, namely from the Cu(I) and Cu(II) oxidation states. From this we can determine which states and excitations to calculate and include in the simulation. More specifically, as detailed in \cite{Gaur2012}, peak A corresponds to a 1s $\rightarrow$ 3d excitation of Cu(II) present in the sample. This then shows in the emission spectrum as a 2p $\rightarrow$ 1s decay with a 3d spectator electron, i.e. a K$\alpha_{1,2}$ transition. A similar effect is described for the B and C peaks, where a 1s $\rightarrow$ 4p excitation in Cu(I) is considered. In this case, a strong solid state effect is considered to explain the existence of two peaks, separated by $\approx$11.2 eV \cite{Gaur2012} in CuO, which is in agreement with the separation of the B and C peaks obtained from the fit (Table \ref{tab:FitResults}) of $10\pm2$.

As described in \cite{Gaur2012}, a splitting of the 4p orbital happens due to the molecular orbital hybridization, where the x and y components of the 4p orbital are coordinated ligand orbitals and the z component is an antibonding orbital. This antibonding nature corresponds to a strong repulsive interaction, which leads to a much higher energy for the C peak. This also means that, statistically, peak C should have half of the intensity of peak B. This is not obvious in Fig. \ref{fig:GalambosiRatios_exc} as the intensity is scaled by the total spectral intensity to match the experimental methodology, however this can be clearly observed by the width of the B and C excitation peaks, where peak C has approximately half of the width of B. The excitations in Cu(I) appear in the spectrum as 2p $\rightarrow$ 1s decays, in this case with a 4p spectator electron. These resonances would not appear in our simulated data, as the respective states and transitions are not included in the simulations if we use the standard way of computing x ray emission spectra of neutral atoms \cite{Guerra2021}.

As the ratios obtained from experiment in the region between $8975$ eV and $8998$ eV have a high uncertainty and not enough data points, we cannot make a definite conclusion regarding the identification of these features, especially taking into account that we are mainly using an atomic model to describe oxides in a metallic matrix. However, by including these excitations, the simulated ratios agree with the ones obtained from experiment in this region. In addition, as the excitation process is a resonant process, this does not impact the ratios at energies higher than $9050$ eV.


As a final remark, all of these discrepancies between the simulated and experimental data could also support the hypothesis that the differences observed in Fig. \ref{fig:GalambosiRatios} and \ref{fig:GalambosiRatios_exc} are due to the disentangling of satellite and diagram intensity in the experimental spectrum as described in Galambosi \textit{et al.} paper \cite{Galambosi2003}. This is particularly significant in the excitation region, where most of what was assumed to be satellite intensity (with spectator holes in the 3d shell) seems to be, in fact, the result of excitation from a different atomic structure (copper oxides with oxidation states of I and II). The differences between simulated and experimental ratios could also be accentuated due to the different theoretical satellite spectrum that was used in our simulations, compared to the one used for the experimental data.
\begin{figure}[h]
    \centering
    \includegraphics[width=\linewidth]{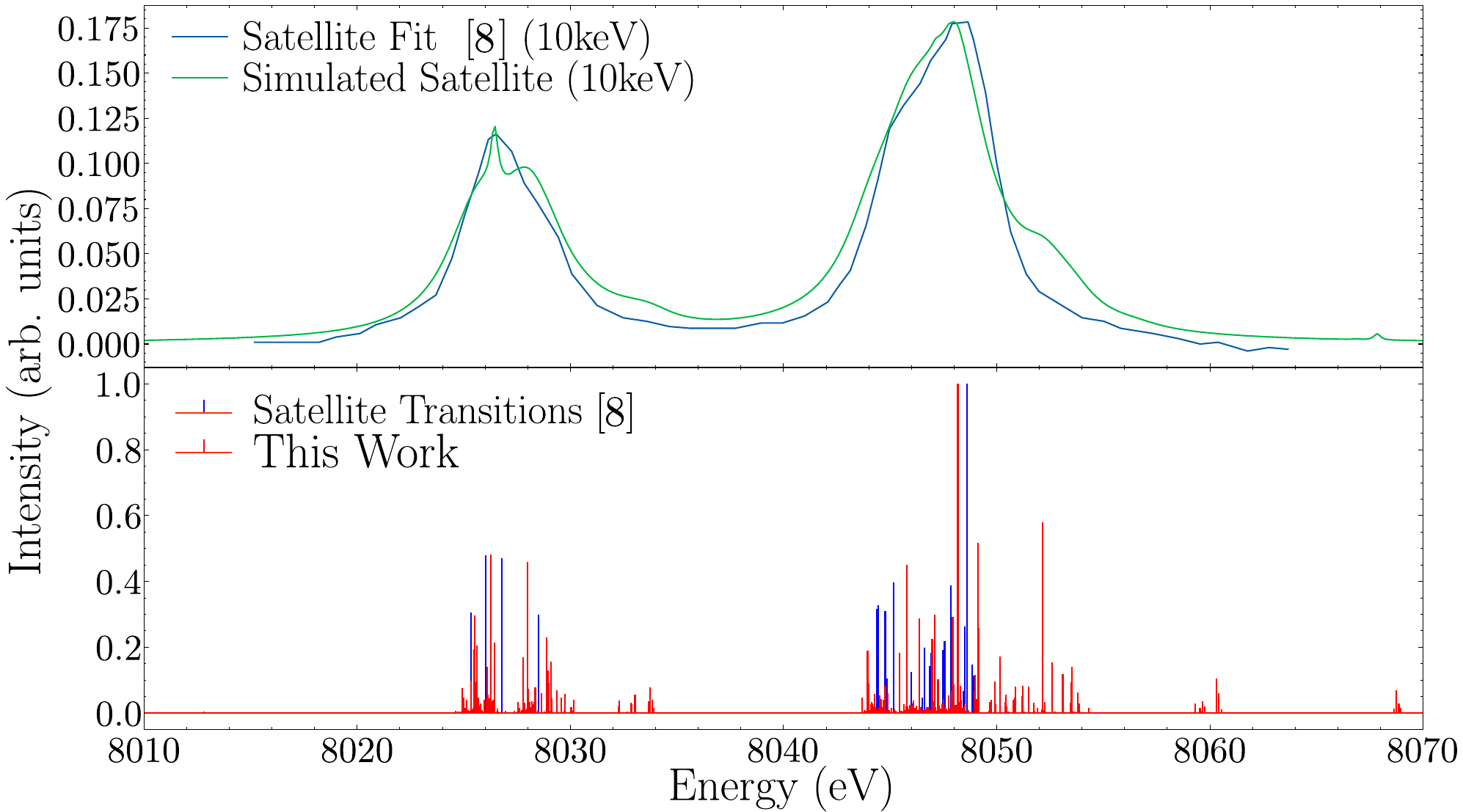}
    \caption{Comparison between the theoretical results for satellite intensity used in \cite{Galambosi2003} and the results presented in this work. The spectrum above shows a simulation for an excitation energy of 10 keV, compared to the same energy experimental fit reported in \cite{Galambosi2003}. Below is a stick graphical representation of the calculated satellite transitions.}
    \label{fig:SatIntensityComp}
\end{figure}
The satellite line shape in Ref. \cite{Galambosi2003} was obtained by Dirac-Fock calculations and fitting, while our was obtained fully \textit{ab initio} from more recent, state-of-the-art Multiconfiguration Dirac-Fock calculations. An example of this can be seen in Fig. \ref{fig:SatIntensityComp}, where we show the spectrum obtained in this work for a beam energy of $10$ keV, compared to the one used in \cite{Galambosi2003} for the same energy. The addition of 3s, 3p and 2p satellites in our calculations change the peak shape, which is more clearly observed by the contributions at $\approx8052$ eV from 3s shake-off, at $\approx8034$ eV from 3p shake-off and at $\approx8060$ from 2p shake-off. The overall shape is also slightly modified by the additional shake-up contributions, in addition to a sharp peak at $\approx8026$ eV which corresponds to 3d shake-up, highlighting the importance of shake-up calculations for high resolution state-of-the-art metrology studies. 


\section{Conclusion} 
The variation of spectral intensity near ionization thresholds is an effect mostly studied using x ray absorption techniques. Nonetheless, this variation can also be studied directly on x ray fluorescence spectra, as we have explored in this work. Several previous experimental studies have been presented where the evolution of spectral intensity is determined from fluorescence spectra, where a theoretical model has also been employed with success.

Although it can be very hard to decompose and quantify each process contributing to experimental fluorescence spectra, which can lead to over- or underestimating some contributions, this becomes trivial for theoretical synthetic spectra.
We have shown that standard state-of-the-art \textit{ab initio} calculations can be used, without any fitting parameter for the intensity from ionizations, simply by using beam parameters and atomic decay data, to predict the mean evolution of spectral intensity near the ionization threshold. This was compared with the models of Thomas and Roy, and a good agreement was obtained for the Thomas model. Given the small uncertainty we assumed for the reported values of Galambosi \textit{et al.}, some of the values for the parameters obtained with the Roy model were not in agreement, nonetheless they were reasonably close to the theoretical values.

In our particular case, the calculations and simulations allowed us to separate the satellite intensity, observed through shake processes, from the direct diagram intensity. Furthermore, we identified a region of the experimental data that could not be described only by ionization data or by the Thomas model. For this, we made further calculations, including excitation data from Cu oxide species, commonly present in solid samples, using a simple atomic model where depending on the oxidation state of the Cu species, we removed one or two of the outer electrons. The decomposition performed with all the data shows a good agreement with experiment, with only small discrepancies. We have attributed these differences to the approximations that were performed in our computation. Namely, the fact that the oxide species were modeled as simple Cu ions, and the disentanglement of satellite and diagram intensities in the experimental work, given that the satellite line shapes that were used by Galambosi \textit{et al.} and ours feature some differences. These differences also highlight the importance of detailed shake calculations, for both shake-off and shake-up, for state-of-the-art metrology studies.

\section*{Acknowledgments}

This research was funded in part by FCT (Portugal) under research center grant UIDB/FIS/04559/2025 (LIBPhys). This work was also funded through the project PTDC/FIS-AQM/31969/2017, “Ultra-high-accuracy x-ray spectroscopy of transition metal oxides and rare earths”. D.P. acknowledges support from FCT through contract 13412/BD/2022 and the project 2022.39784.CPCA.A0. C.G. acknowledges support from FCT, under Contract No. 2022.11197.BD.

\bibliography{Cu_Thomas}

@article{Marques2020,
author = {Marques, Jos{\'{e}} P. and Sampaio, Jorge M. and Santos, Jos{\'{e}} Paulo and Indelicato, Paul and Parente, Fernando},
doi = {10.1002/xrs.3056},
issn = {0049-8246},
journal = {X-Ray Spectrometry},
month = {jan},
number = {1},
pages = {69--73},
title = {{Theoretical fluorescence yields and widths of the K shell and the L subshells, for the Ne, Ar, and Kr isoelectronic sequences}},
volume = {49},
year = {2020}
}

@article{Kochur2006,
   author = {A. G. Kochur and V. A. Popov},
   doi = {10.1088/0953-4075/39/16/016},
   issn = {09534075},
   issue = {16},
   journal = {Journal of Physics B: Atomic, Molecular and Optical Physics},
   month = {8},
   pages = {3335-3344},
   title = {Probabilities of multiple shake processes in sudden approximation},
   volume = {39},
   year = {2006}
}

@article{Galambosi2003,
   author = {S. Galambosi and H. Sutinen and A. Mattila and K. Hämäläinen and R. Sharon and C. C. Kao and M. Deutsch},
   doi = {10.1103/PhysRevA.67.022510},
   issn = {10941622},
   issue = {2},
   journal = {Physical Review A - Atomic, Molecular, and Optical Physics},
   pages = {5},
   title = {Near-threshold multielectronic effects in the Cu $K\alpha_{1,2}$ x-ray spectrum},
   volume = {67},
   year = {2003},
}

@article{thomas1984,
   author = {T Darrah Thomas},
   doi={10.1103/PhysRevLett.52.417},
   issue = {2},
   journal = {PHYSICAL REVIEW LETTERS},
   title = {Transition from Adiabatic to Sudden Excitation of Core Electrons},
   volume = {5},
   year = {1984},
}

@article{stohr1983,
   author = {J. Stöhr and R. Jaeger and J. J. Rehr},
   doi = {10.1103/PhysRevLett.51.821},
   issn = {0031-9007},
   issue = {9},
   journal = {Physical Review Letters},
   month = {8},
   pages = {821-824},
   title = {Transition from Adiabatic to Sudden Core-Electron Excitation: N2 on Ni(100)},
   volume = {51},
   year = {1983},
}

@article{carlson1965,
   author = {Thomas A. Carlson and Manfred O. Krause},
   doi = {10.1103/PhysRev.137.A1655},
   issn = {0031-899X},
   issue = {6A},
   journal = {Physical Review},
   month = {3},
   pages = {A1655-A1662},
   title = {Atomic Readjustment to Vacancies in the  K  and  L  Shells of Argon},
   volume = {137},
   year = {1965},
}

@article{Nguyen2022,
   author = {T. V. B. Nguyen and H. A. Melia and F. I. Janssens and C. T. Chantler},
   doi = {10.1103/PhysRevA.105.022811},
   issn = {2469-9926},
   issue = {2},
   journal = {Physical Review A},
   month = {2},
   pages = {022811},
   title = {Multiconfiguration Dirac-Hartree-Fock theory for copper K$\alpha$ and K$\beta$ diagram lines, satellite spectra, and \textit{ab initio} determination of single and double shake probabilities},
   volume = {105},
   year = {2022},
}

@article{Pinheiro2022,
   author = {Daniel Pinheiro and André Fernandes and César Godinho and Jorge Machado and Gonçalo Baptista and Filipe Grilo and Luís Sustelo and Jorge M. Sampaio and Pedro Amaro and Roberta G. Leitão and José P. Marques and Fernando Parente and Paul Indelicato and Miguel de Avillez and José Paulo Santos and Mauro Guerra},
   doi = {10.1016/j.radphyschem.2022.110594},
   issn = {0969806X},
   journal = {Radiation Physics and Chemistry},
   month = {2},
   pages = {110594},
   publisher = {Elsevier BV},
   title = {K- and L-shell theoretical fluorescence yields for the Fe isonuclear sequence},
   volume = {203},
   year = {2022},
}

@article{Guerra2021,
author = {Guerra, Mauro and Sampaio, Jorge M. and V{\'{i}}lia, Gon{\c{c}}alo Raposo and Godinho, C{\'{e}}sar A. and Pinheiro, Daniel and Amaro, Pedro and Marques, Jos{\'{e}} Pires and Machado, Jorge and Indelicato, Paul and Parente, Fernando and Santos, Jos{\'{e}} Paulo},
doi = {10.3390/atoms9010008},
issn = {2218-2004},
journal = {Atoms},
month = {jan},
number = {1},
pages = {8},
title = {{Fundamental Parameters Related to Selenium K$\alpha$ and K$\beta$ Emission X-ray Spectra}},
volume = {9},
year = {2021}
}

@article{Liu2014,
   author = {Y. P. Liu and C. Gao and J. L. Zeng and J. M. Yuan and J. R. Shi},
   doi = {10.1088/0067-0049/211/2/30},
   issn = {00670049},
   issue = {2},
   journal = {Astrophysical Journal, Supplement Series},
   publisher = {Institute of Physics Publishing},
   title = {Atomic data of Cu i for the investigation of element abundance},
   volume = {211},
   year = {2014}
}

@article{Pradhan2025,
   author = {Anil K. Pradhan and Sultana N. Nahar},
   doi = {10.3390/atoms13100085},
   issn = {22182004},
   issue = {10},
   journal = {Atoms},
   month = {10},
   publisher = {Multidisciplinary Digital Publishing Institute (MDPI)},
   title = {The Opacity Project: R-Matrix Calculations for Opacities of High-Energy-Density Astrophysical and Laboratory Plasmas},
   volume = {13},
   year = {2025}
}

@article{Nahar2023,
   author = {S. N. Nahar and L. Zhao and W. Eissner and A. K. Pradhan},
   month = {8},
   doi = {10.48550/arXiv.2308.14854},
   title = {R-Matrix calculations for opacities.II. Photoionization and oscillator strengths of iron ions FeXVII, FeXVIII and FeXIX},
   year = {2023},
   journal = {},
}

@article{Guerra2017,
   author = {M. Guerra and P. Amaro and J. P. Santos and P. Indelicato},
   doi = {10.1016/j.adt.2017.01.001},
   issn = {10902090},
   journal = {Atomic Data and Nuclear Data Tables},
   month = {9},
   pages = {439-457},
   publisher = {Academic Press Inc.},
   title = {Relativistic calculations of screening parameters and atomic radii of neutral atoms},
   volume = {117-118},
   year = {2017}
}

@article{Guerra2015,
   author = {M. Guerra and J. M. Sampaio and T. I. Madeira and F. Parente and P. Indelicato and J. P. Marques and J. P. Santos and J. Hoszowska and J. Cl Dousse and L. Loperetti and F. Zeeshan and M. Müller and R. Unterumsberger and B. Beckhoff},
   doi = {10.1103/PhysRevA.92.022507},
   issn = {10941622},
   issue = {2},
   journal = {Physical Review A - Atomic, Molecular, and Optical Physics},
   pages = {1-9},
   title = {Theoretical and experimental determination of L -shell decay rates, line widths, and fluorescence yields in Ge},
   volume = {92},
   year = {2015}
}

@article{Wang2025,
   author = {Feilu Wang and Jianrong Shi and Evgeny Stambulchik and Gang Zhao},
   doi = {10.1140/epjd/s10053-025-01032-8},
   issn = {1434-6060},
   issue = {7},
   journal = {The European Physical Journal D},
   month = {7},
   pages = {82},
   title = {Atomic data for modeling of cold photoionized copper},
   volume = {79},
   year = {2025}
}

@article{Chernysheva2024,
   author = {L.V. Chernysheva and V.G. Yarzhemsky},
   doi = {10.1016/j.adt.2024.101650},
   issn = {0092640X},
   journal = {Atomic Data and Nuclear Data Tables},
   month = {7},
   pages = {101650},
   title = {Photoionization cross-sections of valence shells of 3d-elements in VUV-soft X-ray spectral region},
   volume = {158},
   year = {2024}
}

@article{Dawra2022,
   author = {Dishu Dawra and Mayank Dimri and A. K. Singh and Alok K. S. Jha and Rakesh Kumar Pandey and Man Mohan},
   doi = {10.1140/epjd/s10053-022-00362-1},
   issn = {1434-6060},
   issue = {3},
   journal = {The European Physical Journal D},
   month = {3},
   pages = {59},
   title = {Theoretical calculations of the photoionization cross sections for the ground and lowest two excited states of Ni XVIII ion},
   volume = {76},
   year = {2022}
}

@article{Singh2019,
   author = {Avnindra K. Singh and Dishu Dawra and Mayank Dimri and Alok K. S. Jha and Man Mohan},
   doi = {10.1140/epjd/e2019-90464-x},
   issn = {1434-6060},
   issue = {5},
   journal = {The European Physical Journal D},
   month = {5},
   pages = {85},
   title = {Relativistic R-matrix calculations of photoionization cross sections of Cu XVIII},
   volume = {73},
   year = {2019}
}

@article{Sardar2016,
   author = {S. Sardar and M. Bilal and M. A. Bari and R. T. Nazir and A. Hannan and M. Salahuddin and M. H. Nasim},
   doi = {10.1093/mnras/stw350},
   issn = {0035-8711},
   issue = {2},
   journal = {Monthly Notices of the Royal Astronomical Society},
   month = {5},
   pages = {1504-1509},
   title = {Dirac R -matrix calculations of photoionization cross-sections of Ni xiii},
   volume = {458},
   year = {2016}
}

@article{Liu1996,
   author = {Jin-chao Liu},
   doi = {10.1088/0256-307X/13/12/006},
   issn = {0256-307X},
   issue = {12},
   journal = {Chinese Physics Letters},
   month = {12},
   pages = {899-901},
   title = {Spin-Orbit Components of Resonant Satellite Photoionization of Atomic Cu},
   volume = {13},
   year = {1996}
}

@article{Fliflet1976,
   author = {Arne W. Fliflet and Hugh P. Kelly},
   doi = {10.1103/PhysRevA.13.312},
   issn = {0556-2791},
   issue = {1},
   journal = {Physical Review A},
   month = {1},
   pages = {312-317},
   title = {Photoionization of the  3 d  ,  3 p  , and  3 s  subshells of Zn i},
   volume = {13},
   year = {1976}
}

@article{Desclaux1975,
author = {Desclaux, J. P.},
doi = {10.1016/0010-4655(75)90054-5},
issn = {00104655},
journal = {Computer Physics Communications},
number = {1},
pages = {31--45},
title = {{A multiconfiguration relativistic DIRAC-FOCK program}},
volume = {9},
year = {1975}
}

@article{Indelicato1990,
author = {Indelicato, P. and Desclaux, J. P.},
doi = {10.1103/PhysRevA.42.5139},
issn = {1050-2947},
journal = {Physical Review A},
month = {nov},
number = {9},
pages = {5139--5149},
title = {{Multiconfiguration Dirac-Fock calculations of transition energies with QED corrections in three-electron ions}},
volume = {42},
year = {1990}
}

@article{Indelicato2007,
author = {Indelicato, P. and Santos, J. P. and Boucard, S. and Desclaux, J.-P.},
doi = {10.1140/epjd/e2007-00229-y},
issn = {1434-6060},
journal = {The European Physical Journal D},
month = {oct},
number = {1},
pages = {155--170},
primaryClass = {physics},
title = {{QED and relativistic corrections in superheavy elements}},
volume = {45},
year = {2007}
}

@article{Santos2005,
author = {Santos, J. P. and Parente, F. and Boucard, S. and Indelicato, P. and Desclaux, J. P.},
doi = {10.1103/PhysRevA.71.032501},
issn = {1050-2947},
journal = {Physical Review A},
month = {mar},
number = {3},
pages = {032501},
title = {{X-ray energies of circular transitions and electron screening in kaonic atoms}},
volume = {71},
year = {2005}
}

@article{Lowdin1955,
author = {L{\"{o}}wdin, Per Olov},
doi = {10.1103/PhysRev.97.1474},
issn = {0031899X},
journal = {Physical Review},
number = {6},
pages = {1474--1489},
title = {{Quantum theory of many-particle systems. I. Physical interpretations by means of density matrices, natural spin-orbitals, and convergence problems in the method of configurational interaction}},
volume = {97},
year = {1955}
}

@article{Trassinelli2017,
   author = {Martino Trassinelli},
   doi = {10.1016/j.nimb.2017.05.030},
   issn = {0168583X},
   journal = {Nuclear Instruments and Methods in Physics Research, Section B: Beam Interactions with Materials and Atoms},
   month = {10},
   pages = {301-312},
   publisher = {Elsevier B.V.},
   title = {Bayesian data analysis tools for atomic physics},
   volume = {408},
   year = {2017}
}

@inbook{Kavi2023,
   author = {Matjaž Kavčič and Matjaž Žitnik},
   doi = {10.1107/S1574870722005535},
   month = {11},
   pages = {653-658},
   title = {Shake-up and shake-off processes},
   year = {2023},
   publisher = {},
}

@inproceedings{Trassinelli2019,
   author = {Martino Trassinelli},
   doi = {10.3390/proceedings2019033014},
   month = {11},
   pages = {14},
   publisher = {MDPI AG},
   title = {The Nested\_fit Data Analysis Program},
   year = {2019}
}

@article{Gaur2012,
   author = {A. Gaur and B. D. Shrivastava},
   doi = {10.12693/APhysPolA.121.647},
   issn = {1898794X},
   issue = {3},
   journal = {Acta Physica Polonica A},
   pages = {647-652},
   publisher = {Polish Academy of Sciences},
   title = {A comparative study of the methods of speciation using X-ray absorption fine structure},
   volume = {121},
   year = {2012}
}

@article{Shimizu2001,
   author = {K. I. Shimizu and H. Maeshima and H. Yoshida and A. Satsuma and T. Hattori},
   doi = {10.1039/b007276l},
   issn = {14639076},
   issue = {5},
   journal = {Physical Chemistry Chemical Physics},
   pages = {862-866},
   title = {Ligand field effect on the chemical shift in XANES spectra of Cu(II) compounds},
   volume = {3},
   year = {2001}
}

@inproceedings{Klysubun2011,
   author = {Wantana Klysubun and Yatima Thongkam and Sorapong Pongkrapan and Krit Won-In and Jiraroj T-Thienprasert and Pisutti Dararutana},
   doi = {10.1007/s00216-010-4219-1},
   issn = {16182642},
   issue = {9},
   booktitle = {Analytical and Bioanalytical Chemistry},
   month = {3},
   pages = {3033-3040},
   title = {XAS study on copper red in ancient glass beads from Thailand},
   volume = {399},
   year = {2011}
}

@article{Alwis2015,
   author = {L. K. H. K. Alwis and Michael R. Mucalo and Bridget Ingham and Peter Kappen},
   doi = {10.1149/2.0321507jes},
   issn = {0013-4651},
   issue = {7},
   journal = {Journal of The Electrochemical Society},
   pages = {H434-H448},
   publisher = {The Electrochemical Society},
   title = { A Combined SNIFTIRS and XANES Study of Electrically Polarized Copper Electrodes in DMSO and DMF Solutions of Cyanate (NCO-), Thiocyanate (NCS-) and Selenocyanate (NCSe-) Ions },
   volume = {162},
   year = {2015}
}

@article{Xu2023,
   author = {Wei Xu and Yujun Zhang and Kenji Ishii and Hiroki Wadati and Yingcai Zhu and Zhiying Guo and Qianshun Diao and Zhen Hong and Haijiao Han and Lidong Zhao},
   doi = {10.3390/condmat8010008},
   issn = {24103896},
   issue = {1},
   journal = {Condensed Matter},
   month = {3},
   publisher = {MDPI},
   title = {Experimental and Theoretical Investigation of High-Resolution X-ray Absorption Spectroscopy (HR-XAS) at the Cu K-Edge for Cu2ZnSnSe4},
   volume = {8},
   year = {2023}
}

@article{Pankin2022,
   author = {I. A. Pankin and A. V. Soldatov},
   doi = {10.1134/S1027451022060192},
   issn = {18197094},
   issue = {6},
   journal = {Journal of Surface Investigation},
   month = {12},
   pages = {934-938},
   publisher = {Pleiades Publishing},
   title = {Pecularities of XANES Simulations for Cu(II) Sites in Square-Planar Coordination in Cu-Exchanged Zeolites},
   volume = {16},
   year = {2022}
}

@misc{Sanson2021,
   author = {Andrea Sanson},
   doi = {10.20517/microstructures.2021.03},
   issn = {27702995},
   issue = {1},
   journal = {Microstructures},
   month = {10},
   publisher = {OAE Publishing Inc.},
   title = {EXAFS spectroscopy: a powerful tool for the study of local vibrational dynamics},
   volume = {1},
   year = {2021}
}

@article{Stern1975,
   author = {E A Stern and D E Sayers and F W Lytle},
   journal = {NUMBER},
   title = {Extended x-ray-absorption fine-structure technique. III. Determination of physical parameters},
   doi = {https://doi.org/10.1103/PhysRevB.11.4836},
   volume = {11},
   year = {1975}
}

@article{Beccara2003,
   author = {S. Beccara and G. Dalba and P. Fornasini and R. Grisenti and F. Pederiva and A. Sanson and D. Diop and F. Rocca},
   doi = {10.1103/PhysRevB.68.140301},
   issn = {1550235X},
   issue = {14},
   journal = {Physical Review B - Condensed Matter and Materials Physics},
   title = {Local thermal expansion in copper: Extended x-ray-absorption fine-structure measurements and path-integral Monte Carlo calculations},
   volume = {68},
   year = {2003}
}

@article{Fornasini2004,
   author = {P. Fornasini and S. A Beccara and G. Dalba and R. Grisenti and A. Sanson and M. Vaccari and F. Rocca},
   doi = {10.1103/PhysRevB.70.174301},
   issn = {01631829},
   issue = {17},
   journal = {Physical Review B - Condensed Matter and Materials Physics},
   month = {11},
   pages = {1-12},
   title = {Extended x-ray-absorption fine-structure measurements of copper: Local dynamics, anharmonicity, and thermal expansion},
   volume = {70},
   year = {2004}
}

@article{Carlson1973,
   author = {Thomas A. Carlson and C. W. Nestor},
   doi = {10.1103/PhysRevA.8.2887},
   issn = {0556-2791},
   issue = {6},
   journal = {Physical Review A},
   month = {12},
   pages = {2887-2894},
   title = {Calculation of Electron Shake-Off Probabilities as the Result of X-Ray Photoionization of the Rare Gases},
   volume = {8},
   year = {1973}
}

@article{Chantler2010,
   author = {C. T. Chantler and J. A. Lowe and I. P. Grant},
   doi = {10.1103/PhysRevA.82.052505},
   issn = {10502947},
   issue = {5},
   journal = {Physical Review A - Atomic, Molecular, and Optical Physics},
   title = {Multiconfiguration Dirac-Fock calculations in open-shell atoms: Convergence methods and satellite spectra of the copper K$\alpha$ photoemission spectrum},
   volume = {82},
   year = {2010}
}

@article{Deutsch1995,
   author = {M Deutsch and G Holzer and J Hartwigs and J Wolf and M Fritsch and E Forster},
   issue = {1},
   journal = {PHYSICAL REVIEW A},
   title = {K$\alpha$ and K$\beta$ x-ray emission spectra of copper},
   doi = {https://doi.org/10.1103/PhysRevA.51.283},
   volume = {51},
   year = {1995}
}

@article{Mukoyama1987,
   author = {Takeshi Mukoyama and Kazuo Taniguchi},
   issue = {2},
   journal = {PHYSICAL REVIEW A},
   title = {Atomic excitation as the result of inner-shell vacancy production},
   volume = {36},
   doi = {https://doi.org/10.1103/PhysRevA.36.693},
   year = {1987}
}

@article{Sauder1977,
   author = {William C Sauder and James R Huddle and J D Wilson2 and Robert E Lavilla},
   issue = {3},
   journal = {PHYSICS LETTERS},
   title = {DETECTION OF MULTIPLET STRUCTURE IN Cu K$\alpha$ 1,2 BY MEANS OF A MONOLITHIC DOUBLE CRYSTAL SPECTROMETER},
   doi = {https://doi.org/10.1016/0375-9601(77)90914-8},
   volume = {63},
   year = {1977}
}

@article{Ito2006,
   author = {Y. Ito and T. Tochio and H. Oohashi and A. M. Vlaicu},
   doi = {10.1016/j.radphyschem.2005.10.023},
   issn = {0969806X},
   issue = {11 SPEC. ISS.},
   journal = {Radiation Physics and Chemistry},
   pages = {1534-1537},
   publisher = {Elsevier Ltd},
   title = {Contribution of the [1s3d] shake process to K$\alpha$1,2 spectra in 3d elements},
   volume = {75},
   year = {2006}
}

@article{Sier2024,
   author = {Daniel Sier and Jonathan W. Dean and Nicholas T. T. Tran and Tony Kirk and Chanh Q. Tran and J. Frederick W. Mosselmans and Sofia Diaz-Moreno and Christopher T. Chantler},
   doi = {10.1107/S2052252524005165},
   issn = {2052-2525},
   issue = {4},
   journal = {IUCrJ},
   month = {7},
   pages = {620-633},
   publisher = {International Union of Crystallography},
   title = {High-accuracy measurement, advanced theory and analysis of the evolution of satellite transitions in manganese K$\alpha$ using XR-HERFD},
   volume = {11},
   year = {2024}
}

@article{Ito2016,
   author = {Y. Ito and T. Tochio and H. Ohashi and M. Yamashita and S. Fukushima and M. Polasik and K. Słabkowska and Syrocki and E. Szymańska and J. Rzadkiewicz and P. Indelicato and J. P. Marques and M. C. Martins and J. P. Santos and F. Parente},
   doi = {10.1103/PhysRevA.94.042506},
   issn = {24699934},
   issue = {4},
   journal = {Physical Review A},
   pages = {1-11},
   title = {K$\alpha$1,2 x-ray linewidths, asymmetry indices, and [KM] shake probabilities in elements Ca to Ge and comparison with theory for Ca, Ti, and Ge},
   volume = {94},
   year = {2016}
}

@article{Ito2015,
   author = {Y. Ito and T. Tochio and S. Fukushima and A. Taborda and J. M. Sampaio and J. P. Marques and F. Parente and P. Indelicato and J. P. Santos},
   doi = {10.1016/j.jqsrt.2014.10.013},
   issn = {00224073},
   journal = {Journal of Quantitative Spectroscopy and Radiative Transfer},
   pages = {295-299},
   publisher = {Elsevier},
   title = {Experimental and theoretical determination of the K$\alpha$2/K$\alpha$1 intensity ratio for zinc},
   volume = {151},
   year = {2015}
}

@article{Ito2018,
   author = {Y. Ito and T. Tochio and M. Yamashita and S. Fukushima and A. M. Vlaicu and Syrocki and K. Słabkowska and E. Weder and M. Polasik and K. Sawicka and P. Indelicato and J. P. Marques and J. M. Sampaio and M. Guerra and J. P. Santos and F. Parente},
   doi = {10.1103/PhysRevA.97.052505},
   issn = {24699934},
   issue = {5},
   journal = {Physical Review A},
   pages = {1-10},
   publisher = {American Physical Society},
   title = {Structure of high-resolution K$\beta$1,3 x-ray emission spectra for the elements from Ca to Ge},
   volume = {97},
   year = {2018}
}

@article{Mendenhall2017,
   author = {Marcus H. Mendenhall and Albert Henins and Lawrence T. Hudson and Csilla I. Szabo and Donald Windover and James P. Cline},
   doi = {10.1088/1361-6455/aa6c4a},
   issn = {13616455},
   issue = {11},
   journal = {Journal of Physics B: Atomic, Molecular and Optical Physics},
   month = {5},
   publisher = {Institute of Physics Publishing},
   title = {High-precision measurement of the x-ray Cu K$\alpha$ spectrum},
   volume = {50},
   year = {2017}
}

@article{Brumboiu2019,
   author = {Iulia Emilia Brumboiu and Olle Eriksson and Patrick Norman},
   doi = {10.1063/1.5083649},
   issn = {00219606},
   issue = {4},
   journal = {Journal of Chemical Physics},
   month = {1},
   pmid = {30709292},
   publisher = {American Institute of Physics Inc.},
   title = {Atomic photoionization cross sections beyond the electric dipole approximation},
   volume = {150},
   year = {2019}
}

@article{Sabbatucci2016,
   author = {Lorenzo Sabbatucci and Francesc Salvat},
   doi = {10.1016/j.radphyschem.2015.10.021},
   issn = {0969806X},
   journal = {Radiation Physics and Chemistry},
   month = {4},
   pages = {122-140},
   title = {Theory and calculation of the atomic photoeffect},
   volume = {121},
   year = {2016}
}

@article{Roy2001,
   author = {M. Roy and J. D. Lindsay and S. Louch and S. J. Gurman},
   doi = {10.1107/S0909049501008160},
   issn = {0909-0495},
   issue = {4},
   journal = {Journal of Synchrotron Radiation},
   month = {7},
   pages = {1103-1108},
   title = {Multiple-electron excitation in X-ray absorption: a simple generic model},
   volume = {8},
   year = {2001}
}

@misc{NISTXRayDatabase,
   author = {R.D. Deslattes and E.G. Kessler Jr. and P. Indelicato and L. de Billy and E. Lindroth and J. Anton and J.S. Coursey and D.J. Schwab and C. Chang and R. Sukumar and K. Olsen and R.A. Dragoset},
   doi = {https://dx.doi.org/10.18434/T4859Z},
   title = {X-ray Transition Energies (version 1.2)},
   year = {2005},
   url = {http://physics.nist.gov/XrayTrans},
   note = {[Accessed 16-03-2026]},
   address = {National Institute of Standards and Technology, Gaithersburg, MD},
}

\end{document}